\documentclass[12pt]{iopart}

\usepackage{graphicx}
\usepackage{subfig}
\usepackage{iopams}
\usepackage{setstack}

\newcommand{\tmix}{\tau_{\rm mix}}

\newcommand{\bx}{\breve{\bi x}}
\newcommand{\hxc}{\langle \hat{\bi x} \rangle_{\rm c}}
\newcommand{\qss}{\Omega_{\sf U}}
\newcommand{\tmax}{\tmix^\star}
\newcommand{\Vss}{V_{\rm ss}}

\begin{document}
\title[The pointer basis and the feedback stabilization of quantum systems]{The pointer basis and the feedback stabilization of quantum systems}

\author{L. Li$^{1,2}$, A. Chia$^{2,3}$, H. M. Wiseman$^2$}

\address{$1$ Key Laboratory of Quantum Information, University of Science and Technology of China, CAS, Hefei 230026, PRC}

\address{$2$ Centre for Quantum Computation and Communication Technology (Australian Research Council); \\
Centre for Quantum Dynamics, Griffith University, Brisbane, Queensland 4111, Australia.}

\address{$3$ Centre for Quantum Technologies, National University of Singapore, Singapore 117543.}

\ead{H.Wiseman@grifﬁth.edu.au}

\begin{abstract}
The dynamics for an open quantum system can be `unravelled' in infinitely many ways, depending on how the environment is monitored, yielding different sorts of conditioned states, evolving stochastically. In the case of ideal monitoring these states are pure, and the set of states for a given monitoring forms a basis (which is overcomplete in general) for the system.  It has been argued elsewhere [D. Atkins \textit{et~al}., Europhys. Lett. {\bf 69}, 163 (2005)] that the `pointer basis' as introduced by Zurek, Habib and Paz  [\PRL {\bf 70}, 1187 (1993)], should be identified with the {\em unravelling}-induced basis which decoheres most slowly. Here we show the applicability of this concept of pointer basis to the problem of state stabilization for quantum systems. In particular we prove that for linear Gaussian quantum systems, if the feedback control is assumed to be strong compared to the decoherence of the pointer basis, then the system can be stabilized in one of the pointer basis states with a fidelity close to one (the infidelity varies inversely with the control strength). Moreover, if the aim of the feedback is to maximize the fidelity of the unconditioned system state with a pure state that is one of its conditioned states, then the optimal unravelling for stabilizing the system in this way is that which induces the pointer basis for the conditioned states. We illustrate these results with a model system: quantum Brownian motion. We show that even if the feedback control strength is comparable to the decoherence, the optimal unravelling still induces a basis very close to the pointer basis. However if the feedback control is weak compared to the decoherence, this is not the case.
\end{abstract}

\noindent{\it Keywords\/}: pointer basis, physically realizable ensemble, feedback control

\maketitle

\section{Introduction}\label{intro}
The interaction between a quantum system and a quantum measurement apparatus, with only unitary evolution, would entangle the two initially uncorrelated systems so that information about the system is recorded in a set of apparatus states \cite{N32}. Because an entangled state exhibits correlations regardless of the system basis in which it is written, this seems to leave an ambiguity about which system observable the apparatus has actually measured. To get around this problem, Zurek noted that a macroscopic apparatus will be continuously interacting with its environment, and introduced the idea of a `pointer basis' for the quantum apparatus \cite{Zur81}. For an ideally engineered apparatus, this can be defined as the set of pure apparatus states which do not evolve and never enter into a superposition \cite{Zur81, ZHP93}. More realistically, the environmental interaction will cause \emph{decoherence}, which turns a quantum superposition of pointer states into a classical mixture, on a time scale faster than that on which any pointer state evolves. In such a context, the original notion has been modified to define the pointer states as the least unstable pure states [3], i.e. the pure states that have the slowest rate of entropy increase for a given coupling to the environment. 

After an apparatus (or, more generally, any quantum system) has undergone decoherence its state will be, in general, mixed. It is represented by a state matrix $\rho$. Mathematically, there are infinitely many ways to write a mixed state  as a convex combination of pure states $\{\pi_k\}_k$ (a basis) with corresponding weights $\{\wp_k\}_k$. We shall refer to the set of ordered pairs $\{ (\wp_k,\pi_k) \}_k$ as a pure-state ensemble. Each ensemble suggests an \emph{ignorance interpretation} for the mixed state: the system is in one of the pure states $\pi_k$, but with incomplete information, one cannot tell which one it is. However, Wiseman and Vaccarro have shown that not all such ensembles are physically equivalent \cite{WV01} --- only some ensembles are `physically realizable' (PR). A PR ensemble $\{ (\wp_k,\pi_k) \}_k$ is one such that an experimenter can find out which pure state out of $\{\pi_k\}_k$ the system is in at all time (in the long time limit), by monitoring the environment to which the system is coupled. Such ensembles exist for all environmental couplings that can be described by a Markovian master equation~\cite{WM10}, and different monitorings result in different `unravellings'~\cite{Car93} of the master equation into stochastic pure-state dynamics. PR ensembles thus make the ignorance interpretation meaningful at all times in the evolution of a single system, as a sufficiently skilled observer could know which state the system is in at any time, without affecting the system evolution.

Zurek's `pointer basis' concept is supposed to explain why we can regard the apparatus as `really' being in one  of pointer states, like a classical object. In other words, it appeals to an ignorance interpretation of a particular decomposition of a mixed state $\rho$ because of the interaction with the environment. But as explained above, the ignorance interpretation does not work for all ensembles; it works only for PR ensembles. It is for this reason that it was proposed in~\cite{ABJW05} that the set of candidate pointer bases should be restricted to the set of PR ensembles. Furthermore, it was shown in~\cite{ABJW05} that different PR ensembles, induced by different unravellings, differ according to the extent in which they possess certain features of classicality. One measure of classicality, which is closely related to that used by Zurek and Paz~\cite{ZHP93}, is the robustness of an unravelling-induced basis against environmental noise. This is the ability of an unravelling to generate a set of pure states $\{ \pi_k \}_k$ with the longest mixing time~\cite{ABJW05}. This is the time it takes for the mixedness (or entropy or impurity) of the initial pure state to increase to some level, on average, when the system evolves unconditionally (i.e.~according to the master equation). Thus it is this set of states that should be regarded as the pointer basis for the system.

In this paper we are concerned with applying these ideas to quantum feedback control~\cite{WM10}. This field has gained tremendous interest recently and already been successfully applied in many experiments~\cite{SCZ11,VMS12,YHN12}. As in classical control, one needs to gain information about the system in order to design a suitable control protocol for driving the system towards a desired state. However, measurements on a quantum system will in general perturb its state while information is being extracted. This back-action of quantum measurements is a key element that sets quantum feedback protocols apart from classical ones and means that one should take additional care in the design of the in-loop measurement.

A class of open systems of special interest are those with linear Heisenberg equations of motion in phase space driven by Gaussian noise. We will refer to these as linear Gaussian (LG) systems. Such systems have received a lot of attention because of their mathematical simplicity and because a great deal of classical linear systems theory can be re-adapted to describe quantum systems~\cite{DHJ+00,WD05}. LG systems arise naturally in quantum optics, describing modes of the electromagnetic field, nanomechanical systems, and weakly excited ensembles of atoms~\cite{WM10}. 

In this paper, we consider using measurement (an unravelling) and linear feedback control to stabilize the state of a LG system to one of the states in the unravelling-induced basis. In particular we show that when the control is strong compared to the decoherence rate (the reciprocal of the mixing time)  of the unravelling-induced basis, the system state can be stabilized with a fidelity close to one. We will show also that choosing the unravelling which induces the pointer basis (as defined above) maximizes the fidelity between the actual controlled state and the target state, for a strong control. Furthermore, we find that even if the feedback control strength is only comparable to the decoherence rate, the optimal unravelling for this purpose still induces a basis very close to the pointer basis. However if the feedback control is weak, this is not the case. 

The rest of this paper is organized as follows. In \sref{PRPB} we formalize the idea of PR ensembles in the context of Markovian evolution by presenting the necessary and sufficient conditions for an ensemble to be PR which were originally derived in~\cite{WV01}. Here we will also define the mixing time which in turn is used to define the pointer basis. In \sref{LGQS} we review LG systems for both unconditional and conditional dynamics. An expression for the mixing time of LG systems will be derived. In \sref{CLGsys}, we add a control input to the LG system and show that it is effective for producing a pointer state. We will take the infidelity of the controlled state as the cost function for our control problem, and show that this can be approximated by a quadratic cost, thus putting our control problem into the class of linear-quadratic-Gaussian (LQG) control problems. Finally in \sref{ExampleQBM} we illustrate our theory for the example of a particle in one dimension undergoing quantum Brownian motion.

\section{Physically realizable ensembles and the pointer basis}\label{PRPB}

\subsection{Steady state dynamics and conditions for physically realizable ensembles}
In this paper we restrict our attention to master equations that describe valid quantum Markovian evolution so that the time derivative of the system state, denoted by  $\dot{\rho}$ has the Lindblad form. This means that there is a Hermitian operator $\hat{H}$ and vector operator $\hat{\bi c}$ such that 
\begin{eqnarray}
\label{Lindblad}
	\dot{\rho} \equiv {\mathcal L} \rho 	           
	                = -i\big[\hat{H},\rho\big] + \hat{\bi c}^\top \rho \hat{\bi c}^\ddag - {1\over 2} \, \hat{\bi c}^\dagger \hat{\bi c} \;\! \rho - \frac{1}{2} \, \rho \;\! \hat{\bi c}^\dagger \hat{\bi c}  \;.
\end{eqnarray}
Note that $\hat{H}$ invariably turns out to be a Hamiltonian or can be interpreted as one. We have defined $\hat{\bi c}^\ddag$ to be the column vector operator
\begin{eqnarray}
\label{TransposeDagger}
	\hat{\bi c}^\ddag \equiv \big( \hat{\bi c}^\dag \big)^\top  \;,
\end{eqnarray}
where $\hat{\bi c}^\dag$ is defined by transposing $\hat{\bi c}$ and then taking the Hermitian conjugate of each element~\cite{CW11a}:
\begin{eqnarray}
	\hat{\bi c}^\dag \equiv \big( \hat{c}^\dag_1, \hat{c}^\dag_2, \ldots, \hat{c}^\dag_l \big) \;.
\end{eqnarray}
We have assumed $\hat{\bi c}$ to be $l \times 1$. This is equivalent to saying that the system has $l$ dissipative channels. For $l=1$ one usually refers to $\hat{c}$ as a Lindblad operator. Similarly we will call $\hat{\bf c}$ a Lindblad vector operator. We will follow the notation in appendix~A of~\cite{CW11a}, and also use the terms environment and bath interchangeably.

Lindblad evolution is, in general, entropy-increasing and will thus lead to a mixed state for the system~\cite{BP02}. Assuming then, the existence of a steady state $\rho_{\rm ss}$, defined by 
\begin{eqnarray}
\label{sss}
{\cal L}\rho_{\rm ss}= 0 \;,
\end{eqnarray} 
we may write
\begin{eqnarray}
\label{SteadyStateEns}
	\rho_{\rm ss} = \sum_{k} \wp_k \, \pi_{k} \;，
\end{eqnarray}
for some ensemble $\{(\wp_k,\pi_k)\}_k$ where each $\pi_k$ is a projector (i.e.~a pure state) and $\wp_k$ is the corresponding probability of finding the system in state $\pi_k$.

\begin{figure}
\centering
\includegraphics[width=0.6\textwidth]{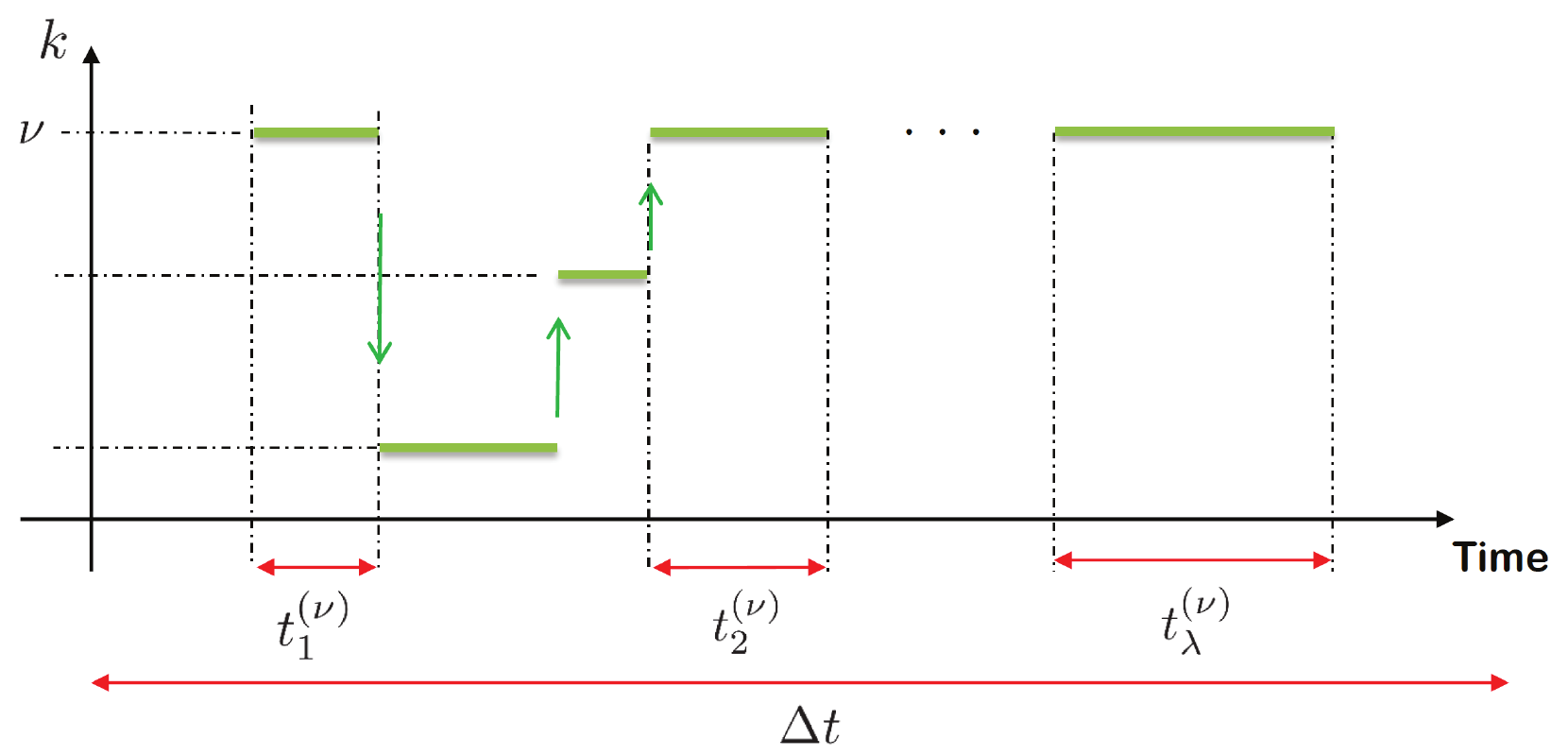} 
\caption{A PR ensemble $\{(\wp_k^{\sf U},\pi^{\sf U}_k)\}$ makes the system state a stationary ergodic process. That is to say the ensemble average over $k$ in \eref{SteadyStateEns} can be obtained by counting, for each value of $k$, the fraction of time the system spends in the $k^{\rm th}$ state for a \emph{single} run of the monitoring $\sf U$ over a sufficiently long period $\Delta t$. The probability of finding the system to be in a particular state, say the state with $k=\nu$ is then $\wp_\nu = \sum_{m=1}^{\lambda} t^{(\nu)}_m / \Delta t$ where for each value of $m$, $t^{(\nu)}_m$ is the amount of time the system spends in state $\pi^{\sf U}_\nu$ before making a jump to a different state as illustrated.}
\label{PRE}
\end{figure}

As explained earlier in \sref{intro}, physical realizability for an ensemble means justifying the ignorance interpretation  of it for all times after the system has reached the steady state. That is, an ensemble is PR if and only if there exists an unravelling ${\sf U}$ (an environmental monitoring scheme which an experimenter can perform) that reveals the system to be in state $\pi^{\sf U}_k$ with probability $\wp_k^{\sf U}$. Note that any ensemble used to represent the system state once it has reached steady state will remain a valid representation thereafter. Thus if the PR ensemble  $\{(\wp_k^{\sf U},\pi^{\sf U}_k)\}$ is to represent $\rho_{\rm ss}$ where the probabilities $\{\wp_k^{\sf U}\}$ are time-independent, then each $\wp_k^{\sf U}$ must reflect the proportion of time that the system spends in the state $\pi_k$. We therefore have a graphical depiction of the system dynamics where it is randomly jumping between the states $\pi^{\sf U}_k$ over some observation interval $\Delta t$. The probability of finding the system to be in state $\pi^{\sf U}_k$ is given by the fraction of time it spends in $\pi^{\sf U}_k$ in the limit of $\Delta t \to \infty$. This is illustrated in~\fref{PRE}. Note that this makes the system state a stationary ergodic process. We now denote the PR ensemble as $\{(\wp_k^{\sf U}, \pi_{k}^{\sf U})\}$, since it depends on the continuous measurement represented by ${\sf U}$. Surprisingly, we can determine whether an ensemble is PR purely algebraically, without ever determining the unravelling $\sf U$ that induces it~\cite{WV01}. Such a method will be employed in \sref{LGQS}.

\subsection{Mixing time and the pointer basis}
\begin{figure}
\centering
\includegraphics[width=0.6\textwidth]{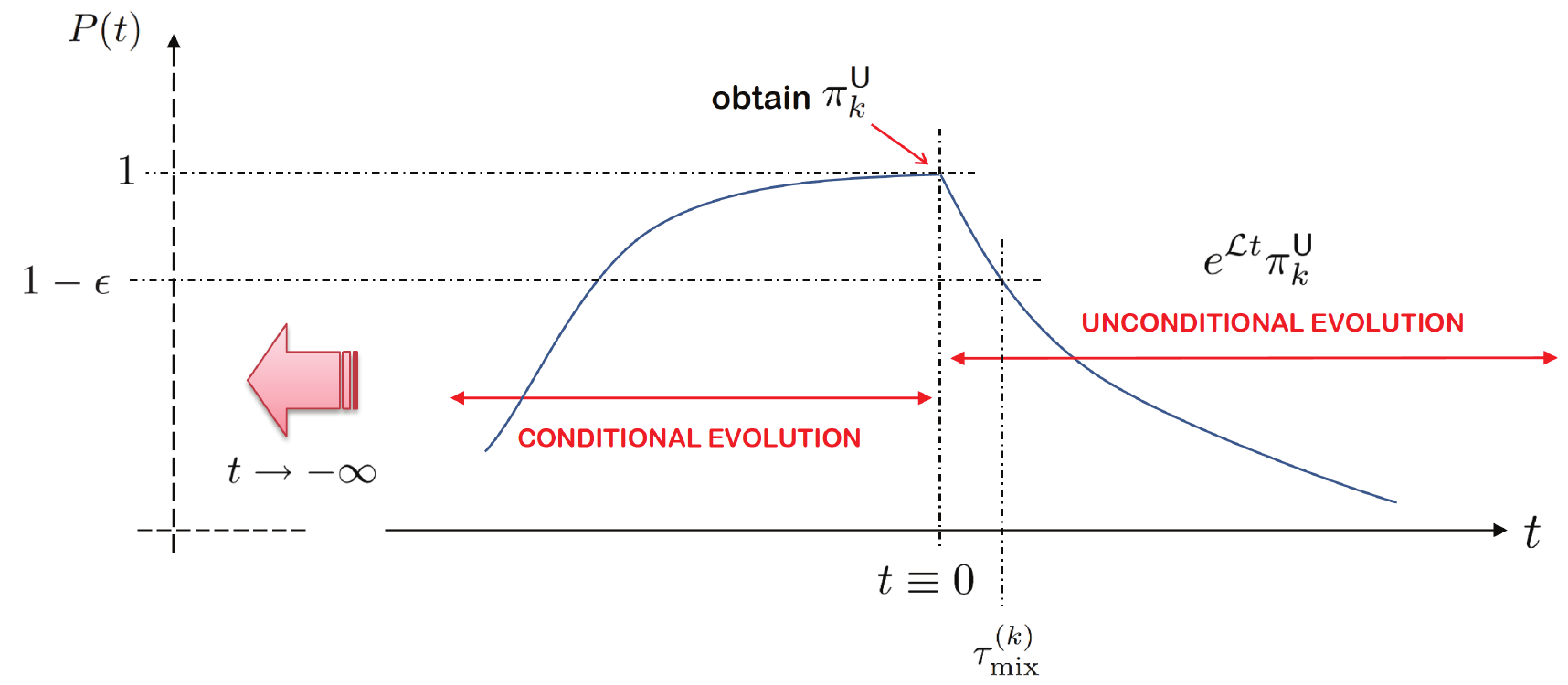} 
\caption{ Illustration of the mixing time for a particular $\pi^{\sf U}_k$ (hence the label $\tmix^{(k)}$ on the time axis). The purity of the system state is denoted by $P(t)$ and we have marked $1-\epsilon$ at a finite distance away from 1 for clarity.}
\label{Tmix}
\end{figure}

The pointer states as defined in \cite{ABJW05} are states which constitute a PR ensemble and, roughly, decohere the slowest. Specifically, Atkins \etal proposed the mixing time $\tau_{\rm mix}$ as the quantity which attains its maximum for the pointer states. This is defined as follows. We assume that an experimenter has been monitoring the environment with some unit-efficiency unravelling ${\sf U}$ for a long (effectively infinite) time so that the conditioned system state is some pure state,  $\pi^{\sf U}_k$. We label this time as the initial time and designate it by $t=0$~\fref{Tmix}). Note the state so obtained belongs to some PR ensemble. The mixing time is defined as the time required on average for the purity to drop from its initial value (being 1) to a value of $1-\epsilon$ if the system were now allowed to evolve unconditionally under the master equation. Thus $\tmix$ is given by the smallest solution to the equation
\begin{equation}
\label{MixingTimeDefn}
{\rm E}\Big\{ {\rm Tr}\Big[ \big\{ \exp( {\cal L} \, \tau_{\rm mix}) \, \pi^{\sf U}_{k} \, \big\}^2 \Big]\Big\}
	= 1 - \epsilon   \;,
\end{equation}
where ${\rm E}\{X\}$ denotes the ensemble average of $X$. Note that \eref{MixingTimeDefn} is a slightly more general definition for the mixing time than the one used in \cite{ABJW05} as $\epsilon$ in \eref{MixingTimeDefn} can be any positive number between 0 and 1. In the next section we will consider the limit of small $\epsilon$.

\section{Linear Gaussian quantum systems}
\label{LGQS}

\subsection{Unconditional dynamics}

A LG system is defined by linear quantum stochastic differential equations driven by Gaussian quantum noise in the  Heisenberg \ picture for (i) the system configuration $\hat{\bi x}$ in phase space, and (ii) the measurement output $\hat{\bi y}$ (also referred to as a current):   
\begin{eqnarray}
\label{LinSys1}
	d \hat{\bi x} & = & A \, \hat{\bi x} \, dt + E \, d \hat{\bi v}_{\rm p}  \;, \\
\label{LinSys2}
	\hat{\bi y}\,\!dt & = & {\sf C} \, \hat{\bi x} \, dt + d \hat{\bi v}_{\rm m} \;.
\end{eqnarray}
Here the phase-space configuration is defined as 2$n$-dimensional vector operator
\begin{equation}
\label{SysConfig}
	\hat{\bi x} \equiv (\hat{q}_1,\hat{p}_1, \hat{q}_2,\hat{p}_2, \ldots,\hat{q}_n,\hat{p}_n)^\top  \;.
\end{equation}
Here $\hat{\bi q} = (\hat{q}_1,\hat{q}_2, \ldots, \hat{q}_n)^\top$ and $\hat{\bi p} = (\hat{p}_1,\hat{p}_2, \ldots, \hat{p}_n)^\top$ represent the canonical position and momentum of the system, defined by
\begin{equation}
   \lfloor \hat{\bi q}, \hat{\bi p} \rceil \equiv \hat{\bi q} \hat{\bi p}^\top - \big( \hat{\bi p} \hat{\bi q}^\top \big)^\top = i \hat{\rm I}_n  \;,
\end{equation}
where $\hbar \equiv 1$ and $\hat{\rm I}_n$ is an $n \times n$ diagonal matrix containing identity operators. All vector operators in \eref{LinSys1} and \eref{LinSys2} are time dependent but we suppressed the time argument, as we will do except when we need to consider quantities at two or more different times.  We take \eref{LinSys1} and \eref{LinSys2} to be It${\rm \hat{o}}$ stochastic differential equations with constant coefficients \cite{Jac10a}, i.e.~$A$, $E$, and ${\sf C}$ are real matrices independent of $\hat{\bi x}$ and time $t$.

The non-commutative nature of $\hat{\bi q}$ and $\hat{\bi p}$ gives rise to the Schr\"{o}dinger-Heisenberg uncertainty relation \cite{Hol11}
\begin{equation}
\label{HeiUncert}
	V + \frac{i}{2} \, Z \ge 0  \;,
\end{equation}
where 
\begin{equation}
\label{DefnOfZ}
	Z \equiv \bigoplus^{n}_{1} \bigg(\begin{array}{cc}
	0 & 1 \\ 
	-1 & 0
	\end{array} \bigg)   \;,
\end{equation}
and $V$ is the covariance matrix of the system configuration, defined by
\begin{equation}
\label{CovarianceDefn}
	V = {\rm Re}\big[ \;\! \big\langle (\hat{\bi x} - \langle \hat{\bi x} \rangle) (\hat{\bi x} - \langle \hat{\bi x} \rangle)^\top \big\rangle \;\! \big]  \;. 
\end{equation}
We are defining the real part of any complex matrix $A$ by ${\rm Re}[A]= (A+A^*)/2$.

The process noise $E \;\! \hat{\bi v}_{\rm p}$ is the unavoidable back-action from coupling the system to the environment. It is a vector operator of Hermitian quantum Wiener increments with a mean and covariance satisfying, for all time,
\begin{eqnarray}
\eqalign
	\langle E \, d \hat{\bi v}_{\rm p}  \rangle = {\bf 0}  \;, \\
\label{ItoProcess}
	{\rm Re} \big[ E \, d\hat{\bi v}_{\rm p} \, d\hat{\bi v}_{\rm p}^\top E^\top \big] \equiv D \, dt \;，
\end{eqnarray}
where for any matrix operator $\hat{\rm A}$ we have defined ${\rm Re} [ \hat{\rm A} ]=(\hat{\rm A}+\hat{\rm A}^\ddagger)/2$ and $\hat{\rm A}^\ddagger$ is defined similarly to \eref{TransposeDagger}. The quantum average is taken with respect to the initial state $\rho(0)$ i.e.~$\langle \hat{\rm A}(t) \rangle = \Tr [\hat{\rm A}(t) \rho(0)]$ since we are in the Heisenberg picture. Note that \eref{ItoProcess} involves the process noise at only one time, second-order moments with $E \;\! d \hat{\bi v}_{\rm p}$ at different times vanish as well as any other higher-order moments. Similarly the measurement noise $d\hat{\bi v}_{\rm m}$ is a vector operator of Hermitian quantum Wiener increments satisfying
\begin{eqnarray}
\eqalign	
	\langle d\hat{\bi v}_{\rm m} \rangle = {\bf 0} \;, \\
\label{ItoMeasurement}
	d\hat{\bi v}_{\rm m} \, d\hat{\bi v}_{\rm m}^\top =  \hat{\rm I}_ R \, dt  \;,
\end{eqnarray}
where we have assumed $d\hat{\bi v}_{\rm m}$ (and also $\hat{\bi y}$) to have $R$ components. As with $E\;\! d\hat{\bi v}_{\rm p}$, \eref{ItoMeasurement} is the only non-vanishing moment for $d\hat{\bi v}_{\rm m}$. The noise $d\hat{\bi v}_{\rm m}$  describes the intrinsic uncertainty in the measurement represented by $\hat{\bi y}$ and in general will be correlated with $E\;\! d\hat{\bi v}_{\rm p}$. We define their correlation by a constant matrix $\Gamma^\top$, i.e.
\begin{equation}
\label{Gamma}
	{\rm Re}\big[ E \, d\hat{\bi v}_{\rm p} \, d\hat{\bi v}_{\rm m}^\top \big] = \Gamma^\top dt \;.
\end{equation}
For the above to describe valid quantum evolution, various inequalities relating $A$, $E$, ${\sf C}$ and $Z$  must be satisfied~\cite{WM10}.
  
Just as a classical Langevin equation corresponds to a Fokker-Planck equation, the quantum Langevin equation \eref{LinSys1} also corresponds to a Fokker-Planck equation for the Wigner function \cite{Sch01} of the system state. Such an evolution equation for the Wigner function can also be derived from the master equation \eref{Lindblad} \cite{Car02}:
\begin{equation}
\label{OUE_Wigner}
\dot{W}( \breve{\bi x} ) = \{- \nabla^{\top}A \breve{\bi x} +\frac{1}{2}\nabla^{\top}D\nabla \} W( \breve{\bi x} ) \;.
\end{equation}
This equation has a Gaussian function as its solution, with mean and covariance matrix obeying 
\begin{eqnarray}
\label{x dynamics}
	 d{\langle \hat{\bi x} \rangle}/dt = A\langle \hat{\bi x} \rangle  \\
\label{V dynamics}
	 d{V}/dt =  A \;\! V + V \;\! A^\top + D \;.
\end{eqnarray}
We restrict to the case that $A$ is Hurwitz; that is, where the real part of each eigenvalue is negative.  Then the steady state Wigner function will be a zero-mean Gaussian \cite{Ris89}
\begin{equation}
\label{Wss}
	W_{\rm ss}( \breve{\bi x } ) = g( \breve{\bi x }; {\bf 0},V_{\rm ss}) \;.
\end{equation}
The notation $\breve{\bi x}$ denotes the realization of the random vector ${\bi x}$ and $g( \breve{\bi x }; {\bi \mu},V)$ denotes a Gaussian with mean ${\bi \mu}$ and covariance $V$ for ${\bi x}$. In this case $V_{\rm ss}$ is the steady-state solution to \eref{V dynamics}; that is, the unique solution of
\begin{equation}	
\label{V steady}
	A \;\! V + V \;\! A^\top + D = 0  \;.
\end{equation}

We saw above that a LG system is defined by the It\^{o} equation \eref{LinSys1} for $\hat{\bi x}$, the statistics of which are characterized by the matrices $A$ and $D$. However, our theory of PR ensembles in \sref{PRPB} was in the Schr\"{o}dinger picture for which the system evolution is given by the master equation \eref{Lindblad}. To apply the idea of PR ensembles to a LG system we thus need to relate $A$ and $D$ to the dynamics specified in the Schr\"{o}dinger picture by ${\cal L}$, which is in turn specified by $\hat{H}$ and $\hat{\bi c}$. One can in fact show that \eref{LinSys1} results from choosing an $\hat{H}$ and $\hat{\bi c}$ that is (respectively) quadratic and linear in $\hat{\bi x}$ \cite{WM10}, i.e.
\begin{equation}
\label{LinSysHamiltonian}
	\hat{H} = \frac{1}{2} \, \hat{\bi x}^\top G \;\! \hat{\bi x}  \;,
\end{equation}
for any $2n \times 2n$ real and symmetric matrix $G$, and 
\begin{equation}
\label{LinSysLindbladOp}
	\hat{\bi c} = \tilde{C} \, \hat{\bi x} \;, 
\end{equation}
where $\tilde{C}$ is $l \times 2n$ and complex. It can then be shown that \eref{LinSysHamiltonian} and \eref{LinSysLindbladOp} leads to 
\begin{eqnarray}
	A & = & Z \big( G + \bar{C}^\top S \bar{C} \big) \label{feeda}; \\
	D & = & Z \bar{C}^\top \bar{C} Z^\top  \label{feedd},
\end{eqnarray}
where we have defined
\begin{eqnarray}
\label{SandCbar}
	S = \left( \begin{array}{cc}
	0 & {\rm I}_l \\ 
	-{\rm I}_l  & 0
	\end{array} \right)\;,  \quad  \bar{C} = \left(  \begin{array}{c}
	{\rm Re}[\tilde{C}]  \\  {\rm Im}[\tilde{C}]
	\end{array}   \right) \;.
\end{eqnarray}
The matrix $S$ has dimensions $2l \times 2l$, formed from $l \times l$ blocks while $\bar{C}$ has dimensions $2l \times 2n$. These definitions will turn out be useful later especially in \sref{ExampleQBM}.

\subsection{Conditional dynamics in the long-time limit}

\Eref{LinSys1} describes only the dynamics of the system due to its interaction with the environment while \eref{LinSys2} describes the dynamics of some bath observable $\hat{\bi y}$ being measured. Our goal in the end is to drive the system to a particular quantum state and this is achieved most effectively if one uses the information obtained from measuring $\hat{\bi y}$. In a continuous measurement of $\hat{\bi y}$ the measurement device will output a continuous stream of numbers over a measurement time $t$. This is typically called a measurement record \cite{JS06} and is defined by
\begin{equation}
\label{MmtRecord}
  {\bi y}_{[0,t)} \equiv \{ {\bi y}(\tau) \, | \, 0 \le \tau < t \} \;, 
\end{equation}
where $\bi{y}(\tau)$ is the result of a measurement of $\hat{\bi y}$ at time $\tau$. In this paper we adopt feedback control in which the controlling signal depends on the ${\bi y}_{[0,t)}$ in~\eref{MmtRecord}. Here we will first explain the system evolution conditioned on knowledge of ${\bi y}_{[0,t)}$ and then from this derive the mixing time using definition \eref{MixingTimeDefn}. The inclusion of a control input in the system dynamics will be covered in~\sref{CLGsys}. 

\begin{figure}
\centering
\includegraphics[width=0.6\textwidth]{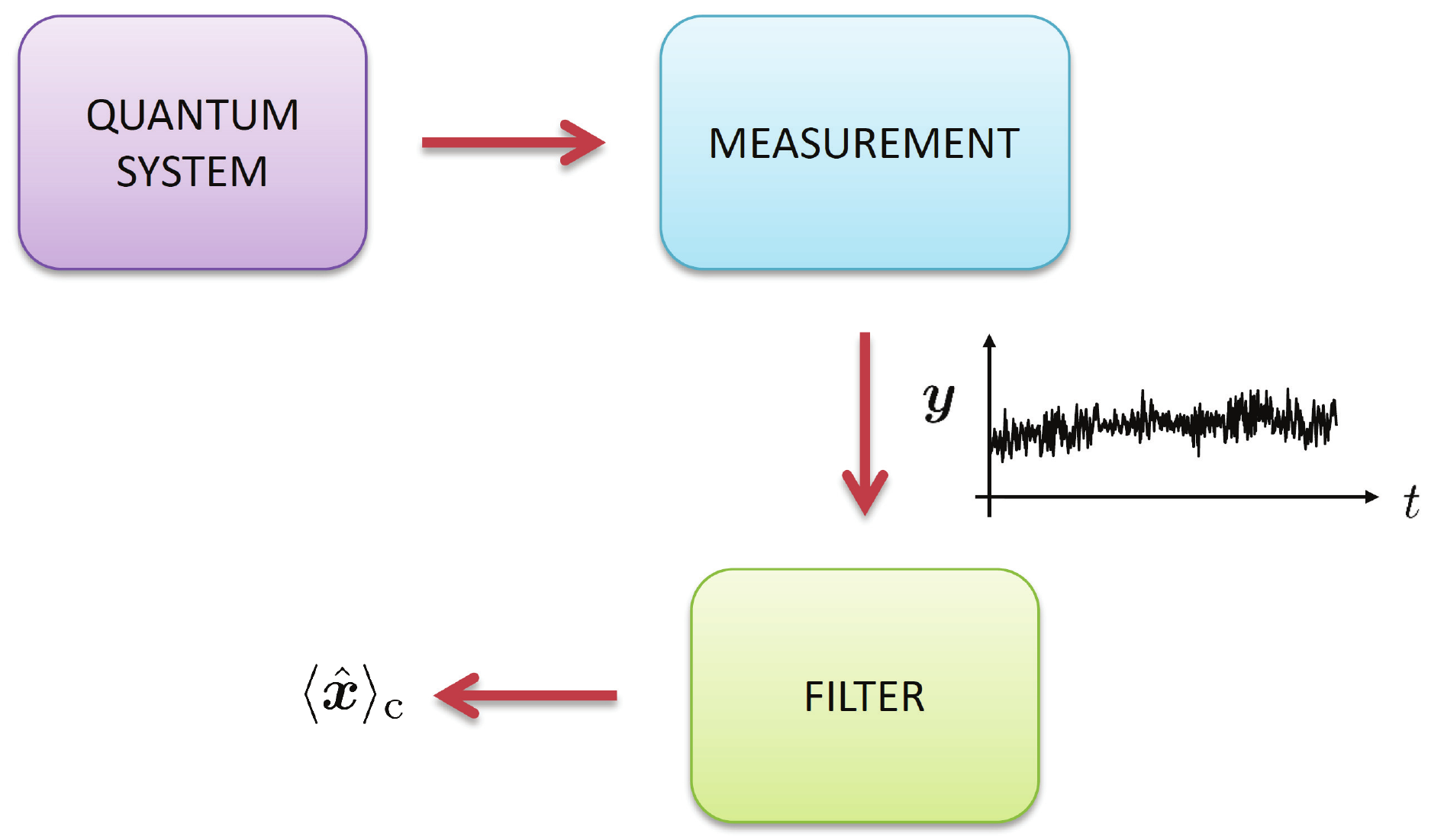} 
\caption{ A filter is a continuous-time estimator which accepts ${\bi y}_{[0,t)}$ as input and produces an estimate of the system configuration as its output. If the mean-square error is used as a performance measure for the filter estimate then the conditional average of $\hat{\bi x}$ is optimal and the filter is characterized by \eref{KB1} and \eref{KB2} in the long-time limit.}
\label{Filter}
\end{figure}

The measured current is first fed into an estimator that uses this information to estimate the system configuration continuously in time. This is often referred to as filtering and the continuous-time estimator is called a filter (see~\fref{Filter}). The performance of the filter may be measured by the mean-square error and it is well known from estimation theory that the optimal estimate is the conditional mean of $\hat{\bi x}$ \cite{KS99}, given by 
\begin{equation}
	\langle \hat{\bi x} \rangle_{\rm c} = \Tr \big[ \hat{\bi x} \;\! \rho_{\rm c}(t) \big]  \;,
\end{equation}
where $\rho_{\rm c}(t)$ is the system state conditioned on ${\bi y}_{[0,t)}$. States as such obey stochastic differential equations that are referred to as quantum trajectories \cite{Car93,Car08} in quantum optics. For control purpose only the evolution of $\langle \hat{\bi x} \rangle_{\rm c}$ matter, and its evolution equation in this case is known as the Kalman-Bucy filter \cite{KB61}. We are ultimately interested in stabilizing the system to some quantum state which, without loss of generality, we can take to have $\langle \hat{\bi x} \rangle_{\rm c} = {\bf 0}$. That is, once the system has reached $\hxc = {\bf 0}$ we would like to keep it there, ideally indefinitely for as long as the feedback loop is running. Thus it is the behaviour of the system in the long-time limit that is of interest to us and it can be shown \cite{WM10} that the Kalman-Bucy filter in this limit is given by 
\begin{equation}
\label{KB1}
	d\hxc = A \, \hxc \, dt + {\rm F}^\top \, d{\bi w}   \;.
\end{equation}
Here $d{\bi w}$ is a vector of Wiener increments known as the innovation \cite{HJS08}, while ${\rm F} \equiv {\sf C} \,\Omega_{\sf U} + \Gamma$, where $\Omega_{\sf U}$ is the solution of the matrix Riccati equation
\begin{equation}
\label{KB2}
A \, \Omega_{\sf U} + \Omega_{\sf U} \, A^\top + D = {\rm F}^\top {\rm F} \;.
\end{equation}
The matrix $\qss$ is the steady-state value of $V_{\rm c}$ [given by \eref{CovarianceDefn} with the averages taken with respect to $\rho_{\rm c}$] and depends on the measurement as indicated by its subscript. It is well known in control theory that when $A$, ${\sf C}$, $E$, and $\Gamma$ [recall \eref{LinSys1},~\eref{LinSys2}, and \eref{Gamma}] have certain properties, $\Omega_{\sf U}$ is a unique solution to \eref{KB2} and is known as a stabilizing solution~\cite{WM10}. We will assume this to be the case in the following theory. As in unconditioned evolution, the conditioned state $\rho_{\rm c}$ also has a Gaussian Wigner function. This is given by
\begin{equation}
\label{Wc}
	W^{\qss}_{\bar{\bi x}}(\bx) = g(\bx;\bar{\bi x},\qss)  \;,
\end{equation}
where we have defined the short-hand $\bar{\bi x}=\hxc$. The uniqueness of $\qss$ means that the conditional states obtained in the long-time limit will all have the same covariance but with different means evolving according to \eref{KB1}. That is, the index $k$ which labels different members of an ensemble representing $\rho_{\rm ss}$ in \eref{SteadyStateEns} is now the vector $\bar{\bi x}$ which changed (continuously) when the system makes `transitions' between different members within an ensemble. Different ensembles are labelled by different values of $\qss$. Such an ensemble is referred to as an uniform Gaussian ensemble.

From \eref{Wss} and \eref{Wc} the ensemble representing the steady state $\rho_{\rm ss}$ of a LG system can be described in terms of Wigner functions as
\begin{equation}
\label{UniformGaussian}
	W_{\rm ss}(\bx) = \int d\bar{\bi x} \; \wp(\bar{\bi x}) \; W^{\qss}_{\bar{\bi x}}(\bx)
\end{equation}
where the distribution of conditional means is another Gaussian, given by
\begin{equation}
\label{P(x)}
	\wp(\bar{\bi x}) = g(\bar{\bi x}; {\bf 0}, V_{\rm ss}-\qss)  \;.
\end{equation}
This can be derived by using \eref{UniformGaussian} to calculate the characteristic function of $\wp(\bar{\bi x})$. 

Since ${\rm F}^\top {\rm F}$ is positive semidefinite by definition, \eref{KB2} implies the linear-matrix inequality for $\qss$:
\begin{equation}
\label{PRconstraint}
	A \;\! \qss + \qss \;\! A^\top + D  \ge 0  \;.
\end{equation}
This constraint together with the Schr\"{o}dinger-Heisenberg relation for the conditional state [i.e.~\eref{HeiUncert} with $V$ replaced by $\qss$]
\begin{equation}
\label{QuantumConstraint}
	\qss + \frac{i}{2} \, Z \ge {} 0 \;,
\end{equation}
are necessary and sufficient conditions for the uniform Gaussian ensemble~\footnote{This can be considered as the \emph{generalized coherent states}~(GCS) for the Heisengberg-Weyl group. See~\cite{KK08}} \eref{UniformGaussian} to be PR \cite{WV01,WM10}. This is the algebraic test for whether an ensemble is PR mentioned in~\sref{PRPB}.

\subsection{Mixing time and the pointer basis}
\label{UforPointerBasis}

As mentioned above, conditioned evolution leads to a Gaussian state with mean $\hxc$ and covariance matrix $\qss$ satisfying \eref{KB1} and \eref{KB2} in the long-time limit. The purity of any Gaussian state with a $2n$-component configuration $\hat{\bi x}$ and covariance $V$ at time $t$ is given by \cite{Oli12}
\begin{equation}
   \label{purity formula}
	P(t) = \frac{1}{\sqrt{ {\rm det}[2V(t)]}} \;,
\end{equation}
where ${\rm det}[A]$ denotes the determinant of an arbitrary matrix $A$. The mixing time [recall \eref{MixingTimeDefn}] is thus defined by 
\begin{equation}
\label{DetVtmix1}
	{\rm det}\big[ 2 V(\tmix) \big] = \frac{1}{(1-\epsilon)^2}   \;,
\end{equation}
where $V(\tmix)$ is the covariance matrix of the state evolved under unconditional evolution from the initial state $\pi^{\sf U}_k$, which has covariance $V(0) = \qss\,$. We have noted in \eref{DetVtmix1} that the ensemble average in \eref{MixingTimeDefn} plays no role since (i) the purity depends only on the covariance; (ii) the different initial states obtained at $t=0$ all have the same covariance $\qss$; and (iii) the evolution of the covariance is independent of the configuration $\hxc$ at all times (not just in steady-state as per \eref{KB2}).

An expression for $\tmix$ can be obtained in the limit $\epsilon \to 0$ by noting that in this limit $\tmix$ will be small so we may Taylor expand $V(t)$ about $t=0$ to first order:
\begin{eqnarray}
\eqalign
	V(\tmix) & = & {} V(0) + \left.\frac{dV}{dt} \right|_{t=0}  \tmix  \\
\label{Vtmix}
	        & = & {} \qss + ( A \, \qss + \qss \, A^\top + D ) \, \tmix  \;.
\end{eqnarray}
Note that we have used~\eref{V dynamics} in \eref{Vtmix}. Multiplying \eref{Vtmix} by $\qss^{-1}$ and taking the determinant gives 
\begin{equation}
\label{DetV}
	{\rm det}\big[2 V(\tmix)\big] = {\rm det}\big[ \, {\rm I}_{2n} + (A \, \qss + \qss \, A^\top + D) \, \qss^{-1} \;\! \tmix \big]  \;,
\end{equation}
where we have noted that the initial state is pure so ${\rm det}\big[2\qss\big] = 1$. For any $n \times n$ matrix $X$ and scalar $\varepsilon$ one can show that
\begin{equation}
\label{DetId}
	{\rm det}\big[ \,{\rm I}_{n} + \varepsilon X \big] \approx 1 + \tr [\,\varepsilon X]   \;,
\end{equation}
for $\varepsilon \to 0\,$. Therefore \eref{DetV} becomes
\begin{equation}
\label{DetVtmix2}
	{\rm det}\big[2V(\tmix)\big] = 1 + \omega \, \tmix  \;,
\end{equation}
where we have defined for ease of writing
\begin{equation}
\label{omegaDefn}
	\omega (\qss) \equiv 2 \, {\rm tr}\big[A\big] + {\rm tr}\big[D\,\qss^{-1}\big]  \;.
\end{equation}
Substituting \eref{DetVtmix2} back into \eref{DetVtmix1} and solving for $\tmix$ we arrive at
\begin{equation}
\label{MixingTime}
	\tmix \approx \frac{2 \epsilon}{\omega}  \;.
\end{equation}
From this expression we see that to maximize the mixing time one should minimize $\omega$. From the definition \eref{omegaDefn} this means that (given $A$ and $D$) $\qss$ should be chosen to minimize $\tr \big[D\,\qss^{-1}\big]$ subject to the constraints \eref{PRconstraint} and \eref{QuantumConstraint}. Since $\qss$ depends on the unravelling $\sf U$, once the $\omega$-minimizing $\qss$ is found, call it $\qss^\star$, we can then find the unravelling that generates $\qss^\star$ by a simple relation \cite{WD05}. The set of pure states that can be obtained by such a measurement therefore forms the pointer basis. In the following we will denote the longest mixing time that is PR by $\tmix^\star$, formally defined by  
\begin{equation}
  \label{tmixStar}
	\tmix^\star \equiv  \frac{2\epsilon} 
	 {\underset{\qss}{\rm min} \; \omega(\qss)}
\end{equation}
subject to  
\begin{eqnarray}
\label{PRCons}
\eqalign
	A \, \qss + \qss \, A^\top + D  \ge  0 \;, \\
\label{QuantumCons}
	          \qss + \frac{i}{2} \, Z  \ge  0 \;, 
\end{eqnarray}
where we have repeated \eref{PRconstraint} and \eref{QuantumConstraint} for convenience. We will denote other quantities associated with $\tmix^\star$ also with a star superscript; in particular, 
\begin{equation}
\label{PointerBasis}
	\qss^\star \equiv \underset{\qss}{\rm arg\;min} \left\{ \big[\omega(\qss)\big] \right\} \;,
\end{equation}
[\;still subject to \eref{PRCons} and \eref{QuantumCons} of course\;] and ${\sf U}^\star$ for the unravelling that realizes the pointer basis. \Eref{PointerBasis} now defines the pointer basis of the system under continuous observation which has a decoherence rate characterized by $1/\tmix^\star\,$. We will illustrate the use of \eref{tmixStar}--\eref{PointerBasis} in~\sref{ExampleQBM} with the example of quantum Brownian motion.

\section{Controlled linear Gaussian quantum systems}
\label{CLGsys}

\begin{figure}
\centering
\includegraphics[width=0.6\textwidth]{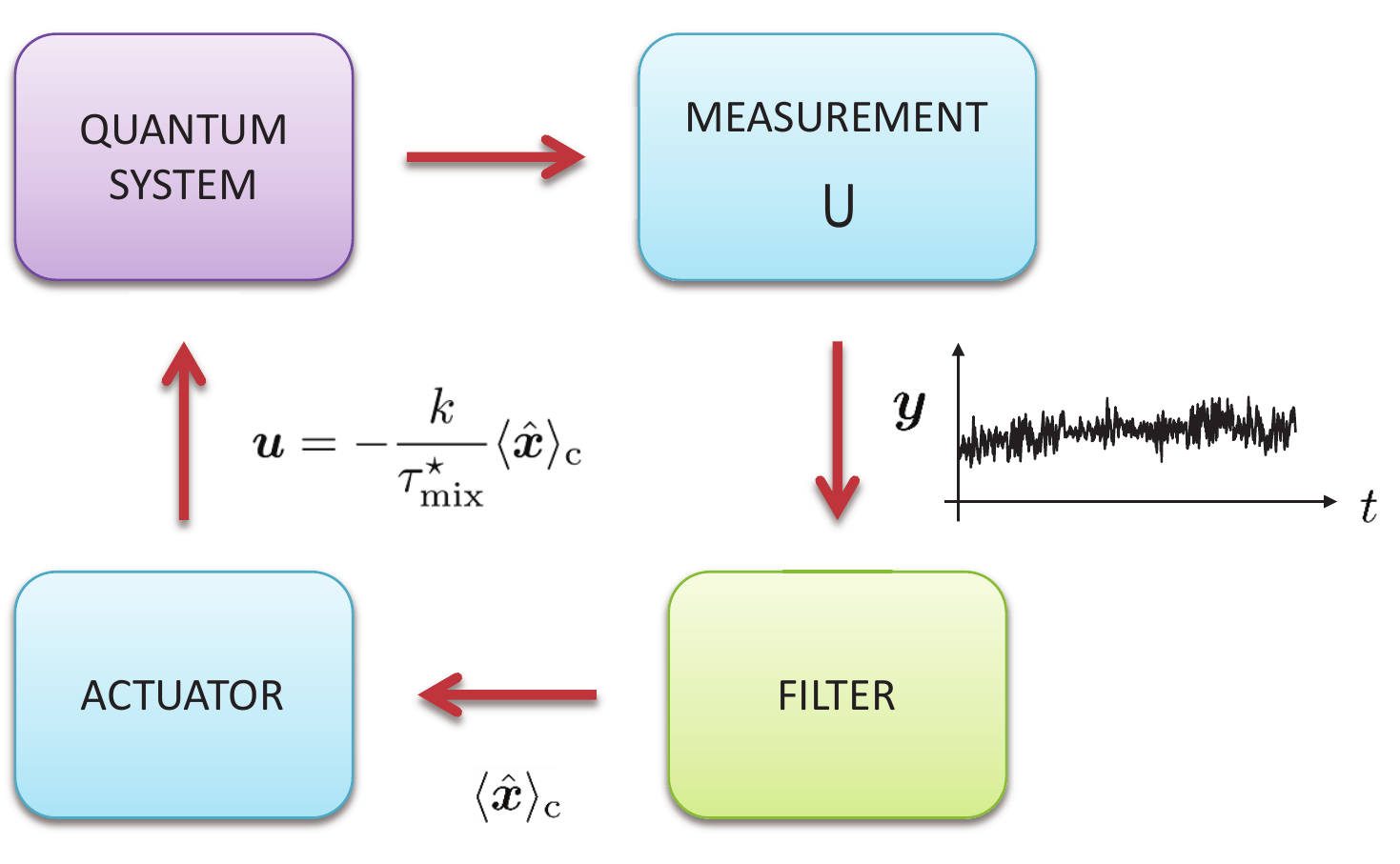} 
\caption{ Feedback loop.}
\label{FeedbackLoop}
\end{figure}

We have said above that for LG systems the unconditioned steady state in phase space is a uniform Gaussian ensemble, where uniformity refers to the fact that each member of the ensemble has the same covariance matrix given by $\qss$. Of the different ensembles the one with $\qss^\star$ identifies the pointer basis and the unravelling ${\sf U}^\star$ that induces it. All that is left to do to put the system into a specific pointer state is to steer the mean of the system configuration $\hxc$  (or, in other words, the centroid of the Wigner distribution in phase space) towards a particular point, say $\hxc = {\bi a}$. This requires feedback control, described by adding a control input ${\bi u}(t)$ that depends on the measurement record ${\bi y}_{[0,t)}$ as shown in~\fref{FeedbackLoop}.   

For simplicity we will define our target state to be at the origin of the phase space, i.e.~${\bi a} = {\bf 0}$. Choosing the phase-space origin will simplify our analysis for a system whose uncontrolled Wigner function does not have a systematic drift away from the origin. This is beneficial for a feedback that is designed to drive the system towards ${\bi a}={\bf{0}}$ simply because the uncontrolled drift does not act against the feedback. In this case one only has to mitigate the effects of diffusion, a process which leads to a greater uncertainty about the system configuration. As this increase in uncertainty can be quantified by the mixing time the effect of the feedback can be characterized by comparing the control strength to $\tmix^\star$. This is illustrated in~\sref{ExampleQBM} using the example of quantum Brownian motion.

\subsection{Adding feedback}

To steer the system towards the origin in phase space we apply a classical input proportional to $\hxc\,$ (effected by the actuator in~\fref{FeedbackLoop})
\begin{equation}
\label{ControlInput}
	{\bi u}(t) = - K \, \hxc(t)  \;,
\end{equation}
where $K$ is a constant matrix, which we take to be
\begin{equation}
\label{FeedbackLG}
	K = \frac{k \epsilon}{\tmax} \: {\rm I}_{2n}  \;.
\end{equation}
Here $k \ge 0$ is a dimensionless parameter which measures the feedback strength relative to the decoherence. This feedback scheme is similar to a special optimal control theory called the linear-quadratic-Gaussian (LQG). In fact, we show in the appendix C that our feedback scheme is equivalent to a limiting case of LQG control. 

The long-time conditional dynamics of the system can thus be written as 
\begin{equation}
\label{KBcontrol}
	d\hxc = N \hxc \, dt + {\rm F}^\top \, d{\bi w} 
\end{equation}
where, 
\begin{equation}
\label{DefnN}
	N \equiv A - K = A -\frac{k \epsilon}{\tmax} \; {\rm I}_{2n}  \;,
\end{equation}
while the equation for the covariance remains unchanged, still given by \eref{KB2}. The control input thus changes only the mean of $\hat{\bi x}$. One can derive from \eref{KBcontrol} the identity
\begin{equation}
\label{RelationM1}
	N \, M + M N^\top + {\rm F}^\top {\rm F} = 0  \;,
\end{equation}
where, as long as $N$ is negative definite,
\begin{equation}
	M \equiv {\rm E_{ss}}\big[ \hxc \hxc^\top \big]  \;,
\end{equation}
with ${\rm E}_{\rm ss}[X]$ denoting the ensemble average of $X$ in the long-time limit (or ``steady state'' \footnote{When referring to $\hxc$ we prefer the term long-time limit as opposed to steady state for the $t \to \infty$ limit since in this limit $\hxc$ still follows a jiggly motion and is not constant as steady state would imply.}). It thus follows that the unconditioned steady state variance matrix in the presence of the feedback is given by
\begin{equation}
\label{RelationM2}
	V_{\rm ss} = \qss + M  \;.
\end{equation}
Relations \eref{RelationM1} and \eref{RelationM2} are useful for calculating the fidelity \cite{NC10,Uhl76} between the controlled state and the target state in the long-time limit.

\subsection{Performance of feedback}
\label{s42}

We take the fidelity between the target state and the state under control to be our performance measure for the feedback loop. The target state has the Wigner function 
\begin{equation}
\label{TargetWigner}
	W_\odot(\bx) = g(\bx;{\bf 0},\qss)  \;,
\end{equation}
while the controlled state is given by
\begin{equation}
\label{ControlledWigner}
	W_{\rm ss}(\bx) = g(\bx;{\bf 0},V_{\rm ss})  \;.
\end{equation}
The fidelity between  states defined by \eref{TargetWigner} and \eref{ControlledWigner} can be shown to be
\begin{equation}
\label{Fidelity}
	F = \frac{1}{\sqrt{{\rm det}\big[ V_{\rm ss} + \qss \big]}}  \;.
\end{equation}
To calculate the determinant in the denominator we note that \eref{RelationM2} gives
\begin{equation}
\label{FidelityDenom}
	\big( \Vss + \qss \big) \, (2\qss)^{-1} =  {\rm I}_{2n} + M \, (2\qss)^{-1}.
\end{equation}
Also note $\det[2\qss] = 1$, we thus have:
\begin{equation}
\label{det[Vss+qss]}
	{\rm det}\big[\Vss + \qss \big] =  {\rm det}\big[ \,{\rm I}_{2n} +  M \qss^{-1}/2 \big] \;.
\end{equation}
To simplify this further we need an expression for $M$, which can be derived by using \eref{RelationM1}. Substituting \eref{KB2} and \eref{DefnN} into \eref{RelationM1} we arrive at  
\begin{equation}
\label{EqnForM}
	M  =  \frac{\tmix^\star}{2k \epsilon} \; \big( A \, \qss + \qss \, A^\top + D \big) + \mathcal{O}((\tmix^\star)^2) \;. 
\end{equation}
Because $\tmix \sim \epsilon$ for  $\epsilon \ll1$,  we may discard second-order terms in $\tmix^\star$ in \eref{EqnForM} to get
\begin{equation}
\label{M(tmix)}
	M \approx \frac{\tmix^\star}{2k\epsilon} \: \big( A \, \qss + \qss \, A^\top + D \big)  \;.
\end{equation}
For strong control ($k \gg 1$), the determinant in \eref{det[Vss+qss]} can then be approximated by an expansion in $M$ to first order. Using \eref{DetId} this gives
\begin{equation}	                               
\label{FidelityDet2}          
	\det \big[\Vss + \qss \big] \approx  1 + \frac{1}{2} \, \tr\big[ M \qss^{-1} \big]   \;.
\end{equation}
Multiplying \eref{M(tmix)} by $\Omega_{\sf U}^{-1}$ on the right and taking the trace we get 
\begin{equation}
\label{TraceMOmegainv}
	\tr \big[ M \qss^{-1} \big] \approx \frac{\omega(\qss)}{2k \epsilon} \: \tmix^\star = \frac{\tmix^\star}{k\tmix} \;.
\end{equation}
Substituting \eref{TraceMOmegainv} into \eref{FidelityDet2} and the resulting expression into the fidelity \eref{Fidelity} we find
\begin{equation}
	F \approx 1 - \frac{1}{4k}\frac{\tmix^\star}{\tmix} \;. 
\end{equation}
That is, the fidelity is close to one for $k$ large (i.e. strong control) as expected. One can also calculate the purity of the feedback-controlled steady state. An expression for this can be obtained from \eref{Fidelity} by replacing $\qss$ by $\Vss$. Then following essentially the same method as for the fidelity calculation we find that it is given by
\begin{equation}
	P \approx 1 - \frac{1}{2k}\frac{\tmix^\star}{\tmix}  \;.
\end{equation}
In both cases we see that the best performance is achieved when $\tmix = \tmix^\star$. That is, when the unravelling generating the most robust ensemble is used. This demonstrates the link between the pointer basis and feedback control for LG quantum systems.

\section{Example: quantum Brownian motion}
\label{ExampleQBM}

We now illustrate the theory of \sref{LGQS} and \sref{CLGsys} with the example of a particle in an environment with temperature $T$  undergoing quantum Brownian motion in one dimension in the high temperature limit. 
This limit means $k_{\rm B} T \gg \hbar \gamma$ where $k_{\rm B}$ is Boltzmann's constant 
and $\gamma$ is the momentum damping rate. In this limit we can use a Lindblad-form master equation as per \eref{Lindblad} to describe the Brownian motion~\cite{DIO93,AB03}, 
with one dissipative channel (i.e.~$l=1$):
\begin{equation}
\label{L2}
  \dot{\rho} = {\cal L} \rho 
                   = - i [\hat{H},\rho] + \hat{c} \rho \hat{c}^\dagger 
                      - \frac{1}{2} \hat{c}^\dagger \hat{c} \rho - \frac{1}{2} \rho \hat{c}^\dagger \hat{c}  \;,
\end{equation}
where
\begin{equation}
	\hat{H} = \frac{\hat{p}^2}{2} + \frac{1}{2} \big( \hat{q} \hat{p} + \hat{p} \hat{q} \big) \;, 
\label{LindbladOperatorQBM}
	\quad  \hat{c} = \sqrt{2T} \hat{q} + \frac{i}{\sqrt{8T}} \; \hat{p}  \;.
\end{equation}
We are using scaled units such that the damping rate, particle mass, Boltzmann constant, and $\hbar$ are all unity. 

The above master equation could also describe a driven and damped single-mode field in an optical cavity with a particular type of optical nonlinearity. In this case $\hat{c}$ is the effective annihilation operator for the field fluctuations (about some mean coherent amplitude). That is, the position $\hat{q}$ and momentum operators $\hat{p}$ of the particle translate into the quadratures of the field mode, with suitable scaling (which depends on the model temperature $T$). This interpretation of the master equation allows the unravellings we discuss below to be easily interpreted: they correspond to homodyne measurement of the cavity output with different local oscillator phases.

Comparing \eref{LindbladOperatorQBM} with \eref{LinSysHamiltonian} and \eref{SandCbar}, we see that $\hat{H}$ and $\hat{c}$ can be written, respectively, as a quadratic and linear function of a two-dimensional configuration defined by 
\begin{equation}
\label{2by2Z}
   \hat{\bi x} = \left( \begin{array}{c}
    \hat{q} \\ \hat{p}
   \end{array} \right)  \;,   \quad  Z = \left( \begin{array}{cc}
       0 & 1 \\ -1 & 0
      \end{array} \right)  \;.
\end{equation}
The matrices $G$ and $\tilde{C}$ in this case are given by
\begin{equation}
\label{2by2c}    
   G = \left( \begin{array}{cc}
       0 & 1 \\ 1 & 1
      \end{array} \right) \;,  \quad   \tilde{C} = \left( \begin{array}{cc}
      \sqrt{2T} , &  i/\sqrt{8T}
      \end{array} \right)  \;.
\end{equation}    
These can then be used to characterize the unconditional dynamics in terms of the drift and diffusion matrices given in \eref{feeda} and \eref{feedd} which are easily shown to be
\begin{equation}
\label{2by2ad}
	A = \left( \begin{array}{cc}
	       0 & 1 \\ 0 & -1
	      \end{array} \right)  \;, \quad D = \left( \begin{array}{cc}
	             1/8T & 0 \\ 0 & 2T
	            \end{array} \right)  \;.
\end{equation}

\subsection{Measurement}
\label{ExampleQBM1}

The theory of PR ensembles, and in particular the realization of a pointer basis by continuous measurement as explained in~\sref{UforPointerBasis} can be applied to the above quantum Brownian motion master equation. 

\begin{figure}[htbp]
	\centering
  	\subfloat[]{
  	\label{pr}
  	\includegraphics[width=0.5\linewidth]{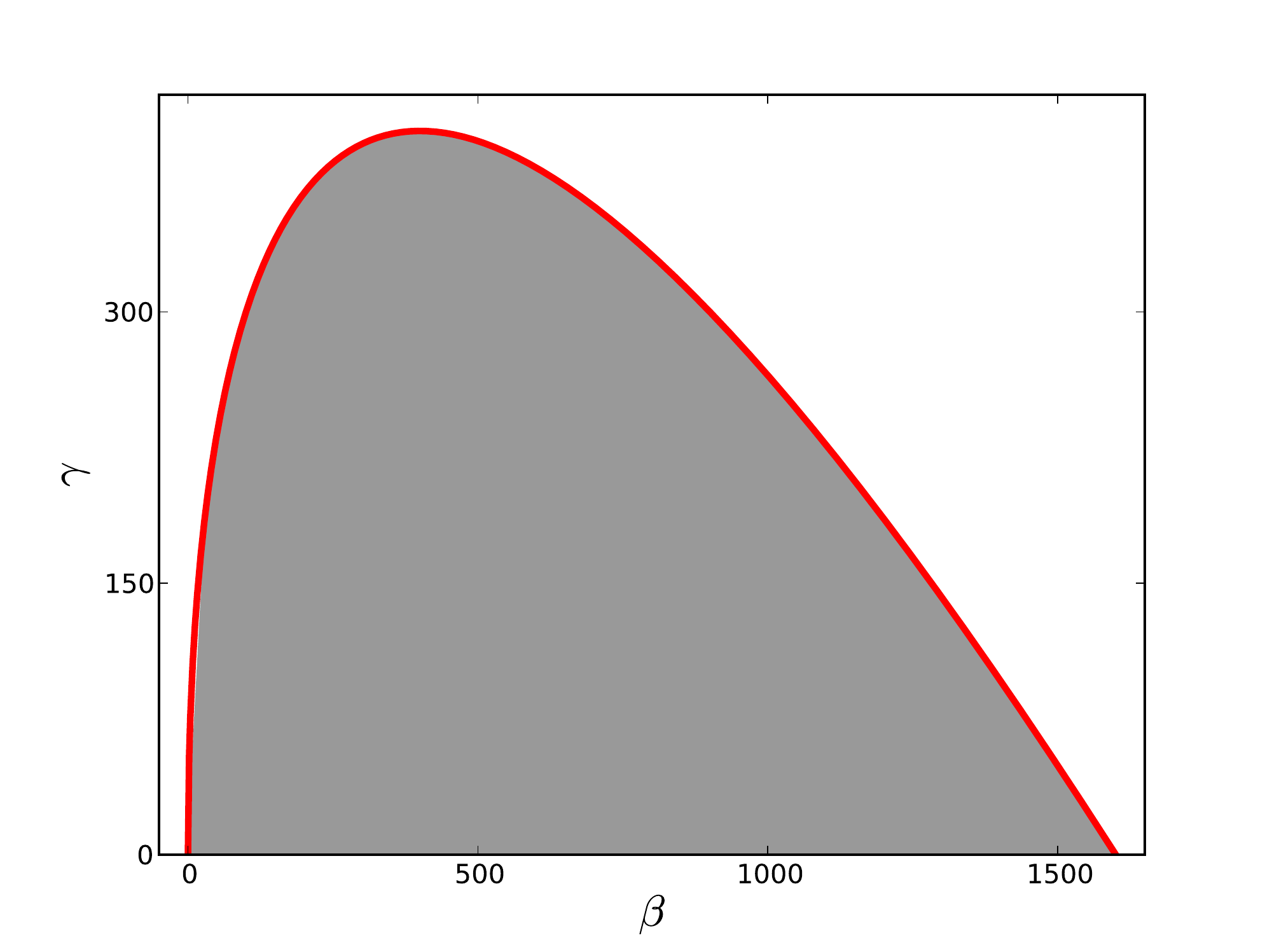}
  	}
   	\hspace{1pt}
	\subfloat[]{
	\label{mix}
	\includegraphics[width=0.4\linewidth]{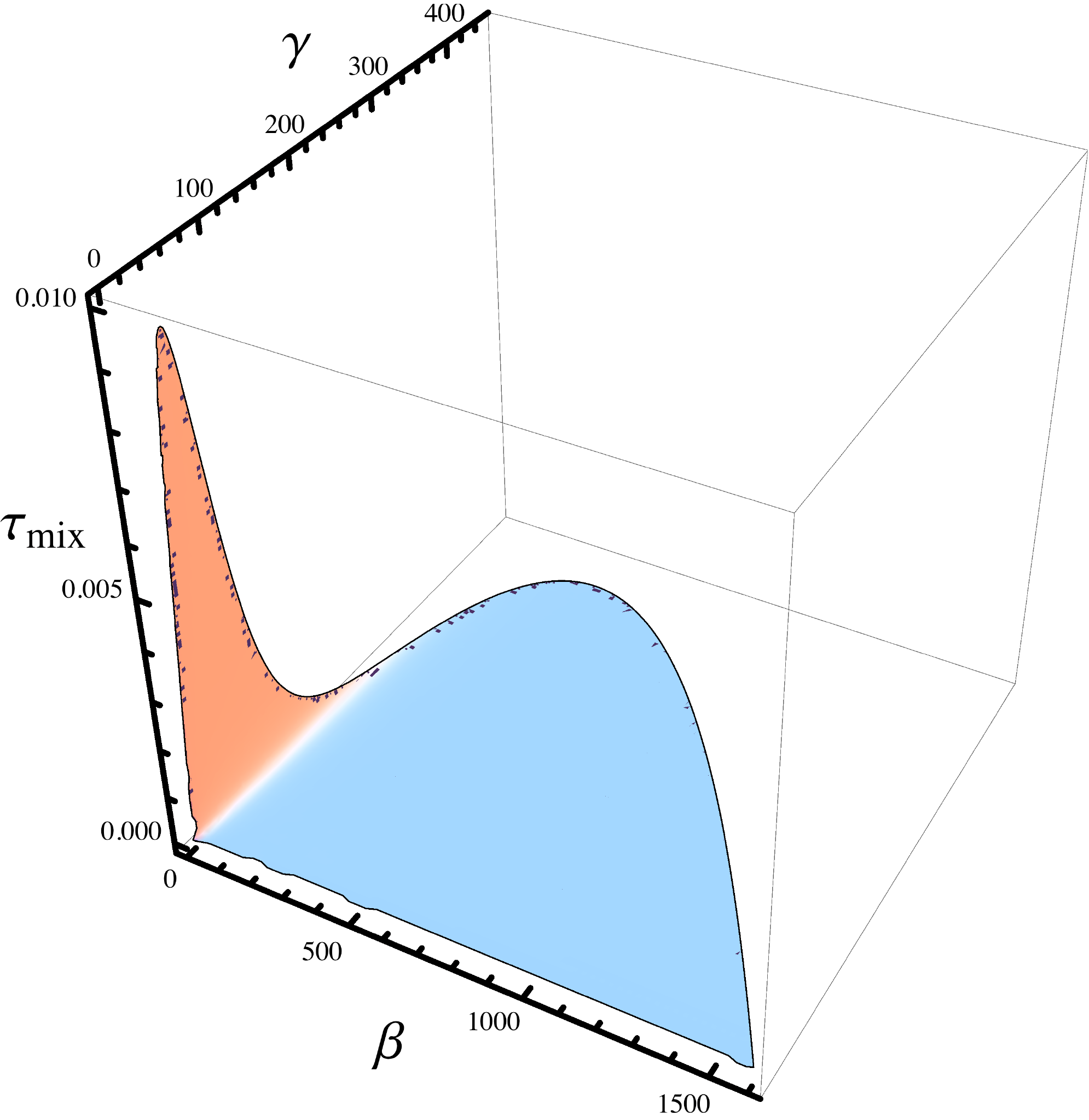}
}
\caption{ Physically realizable region and mixing time for $T$=100. (a) The PR region defined by \eref{QBMPRa} (shaded area). For $T \gg 1$ we find that $0 \leq \gamma \leq 4T$ and $0 \leq \beta \leq 16T$ as can be seen from the plot. (b) The mixing time over the PR region when $\epsilon = 0.1$. It can be seen that the longest mixing time is for ensembles on the boundary of the PR region. That is, the pointer basis lies on the boundary. These plots remain qualitatively the same for all large $T$.}
\end{figure}

Recall that for LG systems with an efficient fixed measurement, the PR ensembles are uniform Gaussian ensembles of pure states, uniform in the sense that every member of the ensemble is characterized by the same covariance matrix $\Omega_{\sf U}$. We showed that for such an ensemble to be a pointer basis,  $\Omega_{\sf U}^\star$ must be the solution to the constrained optimization problem defined by \eref{PRCons}--\eref{PointerBasis}. To find $\Omega^\star_{\sf U}$ let us first write $\qss$ as

\begin{equation}
\label{icm}
	\qss = \frac{1}{4} \left( 
	\begin{array}{cc}
	 \alpha & \beta \\ \beta & \gamma
	\end{array} 
	\right)\;,
\end{equation}
which should satisfy the two linear matrix inequalities \eref{PRCons} and \eref{QuantumCons}. The second of these (the Schr\"{o}dinger-Heisenberg uncertainty relation) is saturated for pure states, and this allows us to write $\alpha$ in terms of $\beta$ and $\gamma$: $\alpha = (\beta^2 +4)/\gamma$. Then from the constraint \eref{PRCons}, we have:
\begin{equation}
\label{QBMPRa}
   \left( \frac{1}{8T}+\frac{\beta}{2} \right) \left( 2T-\frac{\gamma}{2} \right) - \frac{(\gamma-\beta)^2}{16}  \geq  0 \;.
\end{equation}
In the case of $T \gg 1$, a simple calculation from \eref{QBMPRa} shows that the allowed solutions are restricted to $\gamma \in [0,4T)$ and $\beta \in [0,16T]$ (with the maximum range of $\beta$ being when $\gamma = 0$). The PR region is a convex shape in $\beta$-$\gamma$ space as plotted in~\fref{pr}.

Now the definition \eref{DetVtmix1} of the mixing time $\tmix$ is connected with $\beta$ and $\gamma$ by an implicit function (see appendix A):
\begin{equation}
	\det [2V(\tmix, \beta, \gamma)] = 1/(1-\epsilon)^2 \;.
\end{equation}
Searching over the PR region, we can find the longest mixing time $\tau_{\rm mix}^\star$, at the point $(\beta^\star,\gamma^\star)$. This point corresponds to $\qss^\star$, from which we can derive the optimal unravelling matrix ${\sf U}^\star$. It can be shown analytically (appendix A) that $(\beta^\star, \gamma^\star)$ always lies on the boundary of the PR region. Such conditioned states are generated by extremal unravellings ${\sf U}$. Physically (in the language of quantum optics) this corresponds to homodyne detection. Although we will not do so, a relation between ${\sf C}$ in \eref{LinSys2} and ${\sf U}^\star$ may be used to show that ${\sf U}^\star$ does indeed always correspond to homodyne measurement.

A similar conclusion was reached in \cite{ABJW05} but for measurements that maximize the survival time $\tau_{\rm sur}$ and only based on numerics. Note that the survival time (see appendix B) is more general than the mixing time in the sense that it captures any deviation of an unconditionally evolved state from the initially conditioned pure state, not just its decrease in purity. This means that typically $\tau_{\rm sur} \le \tmix$. We show analytically in appendix B that $\tau_{\rm sur}$ is always maximized by PR ensembles that lie on the boundary of the PR region. This result thus rigorously justifies the claim of~\cite{ABJW05} and it is not surprising to find that they maximize $\tmix$ as well (appendix A).

\begin{figure}[htbp]
\includegraphics[width=1.\linewidth]{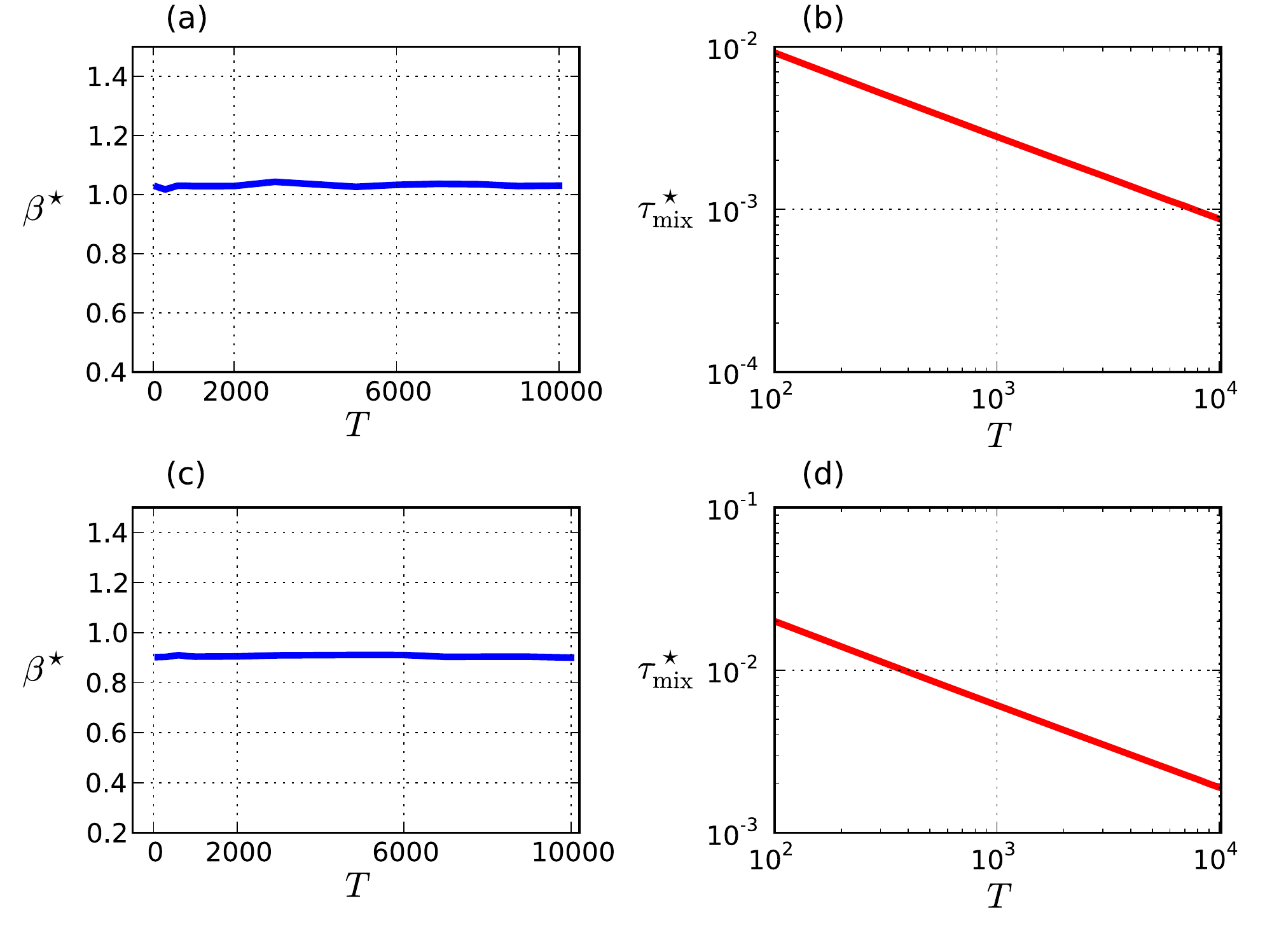}
\caption{ (a), (c): $\beta^\star$ as a function of $T$ for $\epsilon$=0.1 and $\epsilon$=0.2 respectively. (b), (d): log-log plot of $\tmix^\star$ as a function of $T$ for $\epsilon$=0.1 and $\epsilon$=0.2 respectively.}
\label{btt}
\end{figure}

We can see from \fref{btt} (b) and (d) that $\tmix^\star$ decreases monotonically as a function of temperature. Physically this is because a finite-temperature environment tends to introduce thermal fluctuations into the system, making it more mixed. By considering $T$ in the range of $10^2$ to $10^4$ we derive numerically a power law for $\tmix^\star$; see~\fref{btt}. The fits are given in~\tref{t1}, and to a good approximation we have $\tmix^\star \sim T^{-1/2}$. Of course this power law will not hold for $T$ small, but in that regime the high-temperature approximation made in deriving \eref{Lindblad} breaks down. \Fref{btt} also shows that $\beta^\star \approx 1$ is independent of $T$. From the equation for the boundary, it follows that $\gamma^\star \approx 4\sqrt{T}$.
\begin{table}
\centering
\begin{tabular}{c|cc|cl}
\hline
 &\multicolumn{2}{c}{Value} &  \multicolumn{2}{c}{Standard Error} \\
 \cline{2-3} 
 \cline{4-5}
$\epsilon$ & $a$ & $b$  & $a$ & \multicolumn{1}{c}{$b$} \\
\hline
0.1 & -1.02297 & -0.50913 & 0.00325 & 9.63577$\times 10^{-4}$ \\
0.2 &  -0.68442 & -0.50959 & 0.00321 & 9.53113$\times 10^{-4}$ \\
\hline
\end{tabular}
\caption{Fitting results for~\fref{btt} (b) and (d). The fit is given by ${\rm log}\: \tmix^\star = b\,{\rm log}\:T +a$. }
\label{t1}
\end{table}

\subsection{Measurement and feedback}

Having fixed our measurement scheme we are now in position to stabilize the system to a state in phase space prescribed by the Wigner function $W_{\odot}(\bx)=g(\bx; {\bf 0}, \Omega_{\sf U})$. To do so  we simply close the feedback loop by adding a control signal in the form of \eref{ControlInput} and \eref{FeedbackLG}:
\begin{equation}
{\bi u}(t)=-\frac{k \epsilon}{\tmix^{\star}}\langle \hat{\bi x}(t)\rangle_{\rm c} \; ,
\end{equation}
where $k \geq 0$ is a dimensionless parameter determining the strength of control. Under controlled dynamics the drift matrix thus changes from $A$ [specified in \eref{2by2Z}] to $N$ [recall \eref{DefnN}] given by
\begin{equation}
	N = \bigg( 
	\begin{array}{cc}
	-k \epsilon/\tmix^\star & 1  \\ 
	0 & -(1+k \epsilon/\tmix^\star)
	\end{array} 
	\bigg) \;,
\end{equation}
This is an upper-triangular matrix so its eigenvalues $\lambda(N)$ may be read off from the diagonal entries:
\begin{equation}
   \lambda(N) = \big\{ -k\epsilon/\tmix^\star , \, -(1+k\epsilon/\tmix^\star) \big\}  \;.
\end{equation}
Since $k\epsilon$ and $\tmix^\star$ are both greater than zero, $N$ is negative definite (or in the language of control theory, `strictly stable', or `Hurwitz stable') and the conditional steady-state dynamics described by \eref{KBcontrol} will indeed be stabilized to a state with zero mean in the phase-space variables. Note that the uncontrolled dynamics has a drift matrix with eigenvalues given by 
\begin{equation}
   \lambda(A) = \big\{ 0,  \, -1\big\}  \;.
\end{equation}
showing that quantum Brownian motion by itself is only ``marginally stable'' (i.e.~the system configuration will not converge unconditionally to zero owing to the zero eigenvalue). Physically this is because nothing prevents the position of the Brownian particle from diffusing away to infinity. This illustrates a ``stabilizing effect'' of the feedback loop that would not otherwise appear.

\begin{figure}[htbp]
\centering
\includegraphics[width=0.7\linewidth]{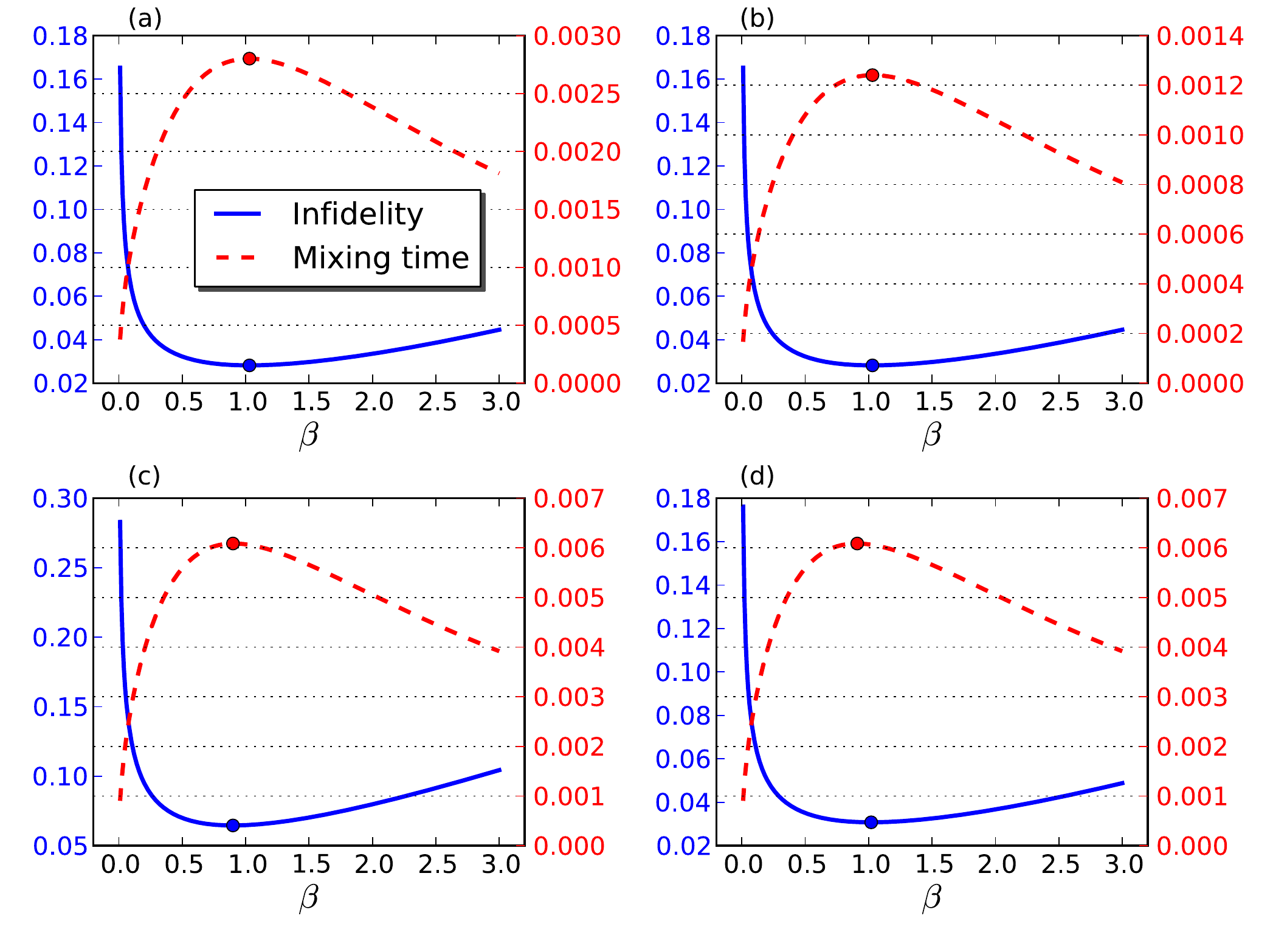}
\caption{ Feedback simulation results. Mixing time (red dash curve) and infidelity (blue curve) for one-dimensional quantum Brownian motion as a function of $\beta$ for (a) $\epsilon = 0.1$, $T=1000$, $k =10$; (b) $\epsilon = 0.1$, $T=5000$, $k=10$; (c) $\epsilon = 0.2$, $T=1000$, $k =5$; and (d) $\epsilon = 0.2$, $T=1000$, $k = 10$. In summary, the effects of changing $T$, $\epsilon$, and $k$ are respectively illustrated in passing from (a) to (b); (a) to (c); and (c) to (d). The left axis stands for the infidelity and the right one stands for the mixing time. The red dot and the blue dot correspond to the maximum mixing time (also corresponds to the $\beta^\star$ point) and the minimal infidelity respectively. See the main text for an explanation of these plots.}
\label{fb1}
\end{figure}

One  may expect that the state of the quantum Brownian particle can be stabilized to the target pointer state \eref{TargetWigner} when the strength of feedback is much greater than the decoherence rate $1/\tmix^\star$. However here we show that the system state can be stabilized to \eref{TargetWigner} very well even when the feedback strength is only comparable to the decoherence rate. This, and the effects of varying $\epsilon$, the environment temperature $T$, and $k$ on the performance of control are depicted in~\fref{fb1} which we now explain.

In~\fref{fb1} we plot the infidelity and the mixing time for $(\beta,\gamma)$ points that saturate the PR constraint \eref{QBMPRa}, as a functions of $\beta$. We do not consider values of $\beta$ and $\gamma$ interior to the PR region as we have already shown that the $\qss^\star$ which generates the pointer basis will lie on the boundary. 

In~\fref{fb1} (a) we set the feedback strength to be comparable to the decoherence rate (corresponding to $k=10$) and for a fixed temperature ($T=1000$). We see from the blue curve in~\fref{fb1}~(a) that the infidelity achieves a minimum close to zero. We also see that our pointer-basis-inducing measurement determined above is indeed optimal for our control objective by observing that the mixing time and the infidelity reaches their maximum and minimum respectively for the same value of $\beta$, namely $\beta^\star$.

To see the effect of the environment temperature we increase $T$ from $1000$ to $5000$ but keep everything else constant. This is shown in~\fref{fb1} (b). As explained previously in~\sref{ExampleQBM1}, an environment at a larger temperature will have a stronger decohering effect on the system and this is seen as the decrease in the mixing time for all values of $\beta$. However, the infidelity, and in particular its minimum value corresponding to $\beta^\star$ has not changed much. This is as expected, since the strength of the feedback is defined relative to the decoherence rate.

Using again~\fref{fb1} (a) for reference we show in~\fref{fb1} (c) the effect of having a larger $\epsilon$ and a smaller $k$ (with $k\epsilon$ fixed). Quantitatively the curves for infidelity and mixing time change, as expected, but qualitatively they are very similar. In particular, the optimal ensemble is at almost the same point for the minimal infidelity, and the value of $\beta^\star$ is little different from that in~\fref{fb1} (a). This is the case even though $\epsilon = 0.2$ barely qualifies as small, and so we would expect some deviations from small $\epsilon$ results obtained in~\sref{s42}.

Finally in~\fref{fb1} (d) we show the effect of increasing the feedback strength by keeping $\epsilon$ and $T$ the same as those in~\fref{fb1} (c) but changing $k$ from 5 back up to 10. As expected this improves the infidelity (i.e.~making it lower for all $\beta$) while the mixing time remains unchanged when compared to that in (c), since it only depends on $\epsilon$ and $T$. We can also compare~\fref{fb1} (d) to (a) which illustrates how the infidelity curve in (d) is restored to one similar to that in (a), as expected because they use the same feedback strength $k$.

\begin{figure}[htbp]
\centering
\includegraphics[width=0.7\linewidth]{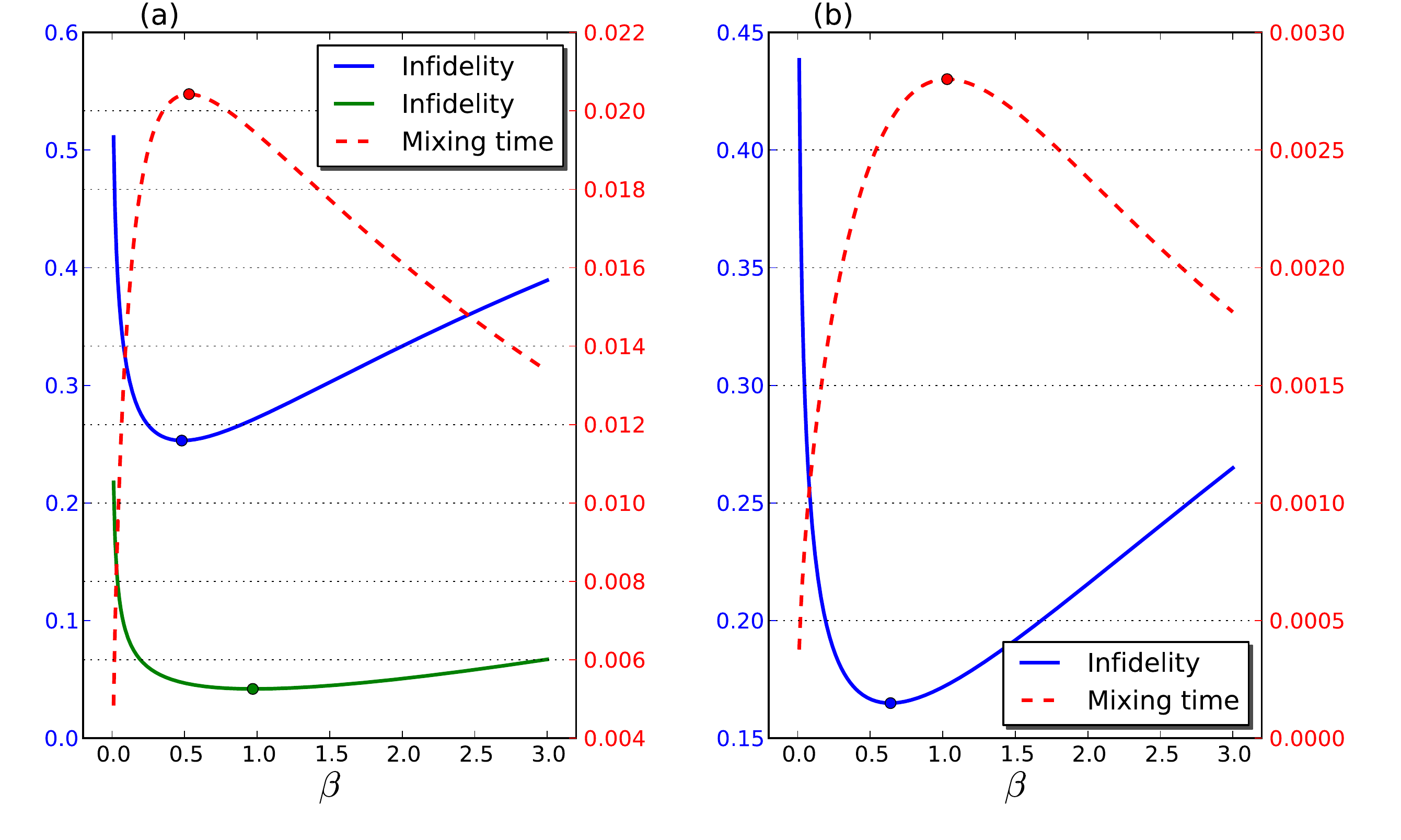}
\caption{ Feedback simulation results. Mixing time (red dash curve) and infidelity (blue and green curve) for one-dimension quantum Brownian motion as a function of $\beta$ for (a) $\epsilon = 0.5$, $T=1000$, $k = 2$ for the blue curve and $k = 10 $ for the green curve; (b) $\epsilon = 0.1$, $T=1000$, $k=2$. The left axis stands for the infidelity and the right one stands for the mixing time. The red dot and the blue (green) dot correspond to the maximum mixing time (also corresponds to the $\beta^\star$ point)  and the minimal infidelity respectively.}
\label{fb2}
\end{figure}

In~\fref{fb2}, we push even further into the regimes where $\epsilon$ is not small, and $k$ is not large. In~\fref{fb2} (a), we choose $\epsilon = 0.5$, and find that the ensemble ($\beta^\star, \gamma^\star$) with the longest mixing time for this threshold of impurity---recall \eref{MixingTimeDefn}---is significantly different from that found with $\epsilon$ small. In the same figure we plot the infidelity of the controlled state with the target state, with $k=10$ and $k=2$. The former (green) gives a minimum infidelity comparable with those with $k=10$ in~\fref{fb1}, and at a similar value of $\beta$. This value of $\beta$ thus differs from the $\beta^\star$ found via maximizing the mixing time. 
This is not surprising as we expect them to be the same only for $\epsilon$ small. The two are closer together, however, for $k=2$ (blue), for which the performance of the feedback is quite poor, as expected. Keeping $k=2$ but restoring $\epsilon$ to a small value of $0.1$ gives somewhat better performance by the feedback control, as shown in~\fref{fb2} (b). 

\section{Conclusion}

We have shown a connection between two hitherto unrelated topics: pointer states and quantum feedback control. While pointer states have appeared in the quantum foundations literature in the early 1980s, the advent of quantum information has since extended this interest in pointer states, and more generally an interest in decoherence into the realm of practical quantum computing \cite{HK97,CMdMFD01,KDV11}. Some of these studies on decoherence have used pointer-state engineering as a means of resisting decoherence such as~\cite{CMdMFD01,KDV11}, but neither work uses feedback \footnote{Note that feedback have been used to protect quantum systems from decoherence as in~\cite{GTV96,HK97}, but not specifically to produce pointer states.}.

Here we have shown that a pointer state, as defined in a rigorous way by us, are those which are most easily attainable, with high fidelity, as target states in quantum linear Gaussian systems. By ``most easily attainable'' we mean with the minimum feedback strength. While we obtained general analytical results in certain limits, our numerical results for a particular system (quantum Brownian motion) shows that our conclusions still hold approximately in a much wider parameter regime. Our work shows how the concept of pointer states has applications outside the realm of quantum foundations, and could aid in the design of feedback loops for quantum LG systems by suggesting the optimal monitoring scheme.

\ack

This research is supported by the ARC Centre of Excellence grant CE110001027. AC acknowledges support from the National Research Foundation and Ministry of Education in Singapore.

\appendix

\section{Robust unravellings for quantum Brownian motion}
\label{ruqbm}
Here we show that the pure-state ensembles on the boundary of the PR constraint are the most robust against decoherence for quantum Brownian motion using mixing time. The maximization of mixing time requires the time dependence of the unconditioned moments so we summarise these first, after which we proceed to consider its optimization.

The Wigner function of one dimensional quantum Brownian motion is of Gaussian form:
\begin{equation}
\label{wignerqbm}
	W_{\bar{p}_t,\bar{q}_t}(p,q,t) = 
	\frac{\exp\{ -\frac{1}{2} (q-\bar{q}_t, p-\bar{p}_t)V_t^{-1}(q-\bar{q}_t, p-\bar{p}_t)^{\top}\}}{2\pi\sqrt{\det (V_t)}},
\end{equation}
where $\bar{q}_t$ and $\bar{p}_t$ are the time dependent mean values of the position and momentum of the particle respectively, and $V_t$ is the covariance matrix:
\begin{equation}
V_t = \left( 
			\begin{array}{cc}
			V_q & V_{qp}  \\ 
			 V_{pq} &  V_p
			\end{array} 
	\right).
\end{equation}
The dynamics in terms of the covariance matrix and the mean values of the Wigner function can then be deduced from \eref{L2} and \eref{LindbladOperatorQBM}:
\begin{eqnarray}
\bar{q}(t)  = \bar{q}_{0}+\bar{p}_{0}(1-e^{-t}), \;\quad \bar{p}(t) = \bar{p}_0 e^{-t}\\
V_{p}(t)  = V_p(0)e^{-2t}+T(1-e^{-2t}), \\\fl
V_{q}(t)  = V_q(0) +\frac{t}{8T} + 2tT + 2[V_{pq}(0)+V_p(0)-2T](1-e^{-t}) \nonumber \\ +[T-V_p(0)](1-e^{-2t}),\\
 \fl V_{pq}(t) =  V_{qp}(t) =  V_{pq}(0)e^{-t}+V_p(0)e^{-t}(1-e^{-t}) + T(1-2e^{-t}-e^{-2t}).
\end{eqnarray}

The mixing time is defined using the purity of a Gaussian state, via equations~\eref{purity formula} and \eref{DetVtmix1}. In the case of quantum Brownian motion, if we take the initial covariance matrix as in~\eref{icm}, the determinant of the time dependent one, defined as $M(t)$, should be a function of $t$, $\beta$ and $\gamma$:
\begin{eqnarray}\fl
M(t, \beta, \gamma)  = \frac{e^{-2t}}{32\gamma T}\left\{ 8(4+ \beta^2)(e^{2t}-1)T^2 
+ \gamma^2[t+8(3-4e^t+e^{2t} + 2t)T^2] \nonumber \right. \\ 
+ 4\gamma T [ 2+ 4\beta(e^t -1)^2 T -32(1-2e^t+e^{2t})T^2 \nonumber\\+ (e^{2t}-1)(1+16T^2)t ]
\left. \right\}
\end{eqnarray}
and we have $M(\tmix, \beta, \gamma) = (1- \epsilon)^{-2}$.

To show the pointer states, which have the longest mixing time, always lie on the boundary of the PR region, we need to prove that for any $\beta$ at a fixed temperature, the mixing time is monotonically increasing with respect to $\gamma$, i.e. $\partial_\gamma \tmix > 0$. First note $\partial_\gamma \tmix = -\partial_{\gamma} M / \partial_{\tmix} M $. Thus we calculate:
\begin{eqnarray}
\partial_{\tmix} M \approx 16 T^2 (4+\beta^2 + 2\beta \gamma \tmix) \label{a1} \;,  \\ 
\partial_\gamma M \approx  -(64+16 \beta^2) T^2 \tmix \;.
\label{a2}
\end{eqnarray}
We have used the fact that $\tmix \ll 1$ above since we are working in the limit of small $\epsilon$. The only variable on the right-hand side of \eref{a1} which can be negative is $\beta$. However, for $T \gg 1$, $\beta$ must take on values between $0$ and $16T$ [see the caption for~\fref{pr}] in which case it is clear that \eref{a1} is positive. Also equation~\eref{a2} is clearly negative. Thus we complete our proof.

\section{Surivival time}
\label{stqbm}

The survival time is defined similarly to the mixing time. It also assumes that a pure state $\pi^{\sf U}_k$ at $t=0$ is obtained as the result of some monitoring represented by ${\sf U}$. We assume that $\pi^{\sf U}_k$ is obtained in the long-time limit with a measurement efficiency of one so that it is pure. Just as in \eref{MixingTimeDefn}, the initial state at $t=0$ is a stochastic quantity since it describes the state of the system after the measurement (which is an inherently random process). The survival time measures how quickly $\pi^{\sf U}_k$ changes on average when allowed to evolve unconditionally. It is defined by
\begin{equation}
\label{surdef}
	{\rm E}\Big\{ {\rm Tr}\Big[ \pi^{\sf U}_k \exp( {\cal L} \, \tau_{\rm sur}) \, \pi^{\sf U}_{k} \Big] \Big\} 
            = 1 - \epsilon  \;.
\end{equation}
As with the mixing time $\epsilon$ here will be assumed to be small. We call the quantity on the left-hand side of \eref{surdef} the survival probability. Note that we do not have to worry about the existence of multiple values of $\tau_{\rm sur}$ in \eref{surdef} because the survival probability can be shown to be a monotonically decreasing function of time for high temperatures (which is the regime that we are working in). We can prove this once we have an expression for the survival probability.

The survival probability can be calculated in phase space where it is expressed as an integral of a product of Wigner functions corresponding to the initial state $\pi^{\sf U}_k$ and its time-evolved version:
\begin{equation}
\label{surprodef}
S_{\bar{p}_0}(\tau) = 2\pi \int_{-\infty}^{+\infty}dp \int_{-\infty}^{+\infty}dq\, W_{\bar{p}_0}(p,q,0)W_{\bar{p}_0}(p,q,\tau).
\end{equation}
Note the absence of the subscript $\bar{q}_0$ in \eref{surprodef}. This is because the overlap is independent of the initial position $\bar{q}_0$. Thus $k$ can be replaced simply by $\bar{p}_0$, and we can set $\bar{q}_0 = 0$ in~\eref{wignerqbm} for ease of calculating the integral in~\eref{surprodef}.

Averaging over the initial states (which are now just labelled by $\bar{p}_0$) gives the survival probability:
\begin{equation}
S(\tau) = \int_{-\infty}^{+\infty} d\bar{p}_0\wp(\bar{p}_0)S_{\bar{p}_0}(\tau).
\end{equation}
Here $\wp(\bar{p}_0) = g(\bar{p}_0; 0, T-V_{p_0})$ is the probability distribution of the initial particle momentum. Writing the initial covariance matrix as in \eref{icm}, and using the time-dependent solution given in appendix A, a rather lengthy calculation shows the survival probability to be given by 
\begin{equation}
\label{sdef}
S(\tau,\gamma,\beta,T) = 4\sqrt{R(\tau, \gamma, T) / G(\tau,\gamma,\beta,T)}
\end{equation}
where $R(\tau,\gamma,T) = e^{2\tau}\gamma T $ and
\begin{eqnarray}
\fl
G(\tau, \gamma, \beta, T) =  -T\gamma\left( \gamma + \beta - 8T \right)^2 + e^{\tau}\left[ 64\left(\tau-2\right)T^{3} \gamma 
- 16T^{2} \left(4+\beta^{2}+2\beta \gamma+(1 - \tau)\gamma^{2}\right) \right.   \nonumber \\
+ 2T\gamma(8-2\tau+\beta^{2}+\gamma^{2}+2\gamma \beta )+\tau\gamma^{2} \big] \nonumber \\
+e^{2\tau}T\bigg[64(\tau-1)T^{2}\gamma
+ 16T(4+\beta ^{2}+\beta \gamma )-\gamma \left[(\beta +\gamma )^{2}-4\tau\right] \bigg] \;.
\end{eqnarray}

To show that the pure-state ensembles which maximize $\tau_{\rm sur}$ are those that lie on the boundary of the PR constraint we note that these ensembles must also maximize the survival probability at a fixed time and a fixed temperature. We can do this by the same technique as in appendix A, using $\partial_{\gamma}\tau = - \partial_{\gamma}S / \partial_{\tau} S$. First, we show that $\partial_{\tau}  S < 0$, as mentioned under \eref{surdef}. 
From~\eref{sdef}, we get:
\begin{equation}
\partial_{\tau}S = \frac{8(G\partial_\tau R - R\partial_{\tau}G)}{SG^2}.
\end{equation}
We thus require the numerator to be negative. As we have done to equations~\eref{a1} and~\eref{a2} ($\tau_{\rm sur} < \tmix \ll 1$ \footnote{In fact, for $T > 1000$ it is still the case that $\tau_{\rm sur} \ll 1$ even for $\epsilon=0.5$}), we can show
\begin{eqnarray}
\label{ds/dg}
G\partial_\tau R - R\partial_{\tau}G  \approx &&-e^{2\tau}\gamma T \left\{\gamma^2 +64\gamma \tau^2 T^3 \right.\nonumber\\ 
&& + 8T^2[2\beta \gamma \tau + \beta^2(2+\tau) + \tau(4+\gamma^2 +2\tau)] \left. \right\}\;.
\end{eqnarray}
For the same reason as in equation~\eref{a1}, it is clear that \eref{ds/dg} is negative. Next we show $\partial_{\gamma}S > 0$. First note
\begin{equation}
\partial_{\gamma}S = \frac{8(G\partial_{\gamma} R - R\partial_{\gamma}G)}{SG^2} \;.
\end{equation}
Then, using the same approximation as above, we get:
\begin{equation}
\label{ds/dt}
G\partial_{\gamma} R - R\partial_{\gamma}G \approx 16 T^4 e^{2\tau} \big[ 4 + \beta^2 + \gamma (1 - \tau) \big]  \;
\end{equation}
which is clearly positive. This concludes our proof that the most robust ensemble, as measured by survival time as well as mixing time, lies on the boundary of the PR region. 

\section{Equivalence to LQG control}
\label{lqgqbm}

The LQG protocol was originally designed for classical systems but it has become widely known in quantum feedback~\cite{DJ99,DHJ+00}. Here we show that our feedback scheme is equivalent to a limiting case of LQG control. To prove this equivalence we first need to see how the LQG problem is defined. This is simple since it also restricts the system to be linear and Gaussian as defined above, except now the control input ${\bi u}(t)$ is required to minimize a quadratic cost function. This is given by, for some starting time $t_0$ and terminal time $t_1$,
\begin{equation}
\label{CostLQG}
	j = \int^{t_1}_{t_0} dt' \Big\{ \langle \hat{\bi x}^\top \;\! P \, \hat{\bi x} \rangle (t') + {\rm E}\big[{\bi u}^\top\!(t') \;\! Q \;\! {\bi u}(t')\big] \Big\}  \;,
\end{equation}
where $P$ and $Q$ are real, time-independent, and symmetric matrices. The matrix $P$ is assumed to be positive semidefinite while $Q$ is positive definite. This expression defines $P$ as the penalty imposed for not making the target configuration stay close to the origin while $Q$ represents the cost of control. Note also that the integrand of \eref{CostLQG} involves only ensemble averages so that $j$ is deterministic. A general expression for the optimal control is well known. This gives a time-dependent ${\bi u}(t)$ which minimizes the cost  \eref{CostLQG} and is therefore more general than what we require. Here we are only interested in the long-time limit. In this case the optimal control has the form of \eref{ControlInput} but $K$ is now given by
\begin{equation}
\label{FeedbackLQG}
	K = Q^{-1} \, Y  \;,
\end{equation}
where $Y$ is the solution of
\begin{equation}
\label{RicattiEqnForY}
	A^\top Y + Y A + P - Y Q^{-1} \, Y = 0  \;,
\end{equation}
which is known as an algebraic Ricatti equation. 

We can now show that our feedback strategy introduced for LG systems is equivalent to LQG control in the sense that there exists a quadratic cost i.e.~an appropriate choice of $P$ and $Q$, for which the cost-minimizing control will reproduce \eref{ControlInput} with a $K$ defined by \eref{FeedbackLG}. Such a choice of $P$ and $Q$ is given by
\begin{equation}
\label{EquivLQG}
	P = k \, \qss^{-1}  \;, \quad Q = \frac{(\tmix^\star)^2}{k\epsilon^2} \; \qss^{-1}  \;.
\end{equation}
Matrix $P$ corresponds to the cost in the formula of $k\tr[ \qss^{-1}\Vss ]$~\cite{WM10}. Note the purity of the controlled state is approximated by $(\tr[ \qss^{-1}\Vss ])^{-1}$, thus to minimize the cost associated with $P$ is equivalent to maximizing the purity of the feedback-stabilized state, which justifies our choice of $P$. Unlike $P$, there is no physical explanation for our choice of $Q$ except that it reproduces the control input given by~\eref{ControlInput} and~\eref{FeedbackLG}. These matrices now implicitly define $K$ through \eref{FeedbackLQG} and \eref{RicattiEqnForY}. To show that the $K$ so obtained agrees with \eref{FeedbackLG} we first substitute \eref{EquivLQG} into \eref{RicattiEqnForY} to get 
\begin{equation}
\label{EqnForY}
	\epsilon^2 \big( Y \qss \big)^2 = (\tmix^\star)^2 \, {\rm I}_{2n} + \frac{(\tmix^\star)^2}{k} \; A^\top Y + \frac{(\tmix^\star)^2}{k} \; Y A  \;.
\end{equation}
Recall that we have expressed the feedback strength, as a $k$-multiple of the decoherence rate. In the limit of $\epsilon \to 0$ we can expect $\tmix^\star \ll 1$ so that any feedback strength which is at least equal to the decoherence rate will make $k \gg \tmix \gg (\tmix^\star)^2$. In this case we can approximate \eref{EqnForY} by
\begin{equation}
\label{ApproxEqnForY}
	\epsilon^2 \big( Y \qss \big)^2 \approx (\tmix^\star)^2 \, {\rm I}_{2n}  \;,
\end{equation}
from which we get
\begin{equation}
\label{SolnForY}
	Y \approx \tmix^\star \qss^{-1} / \epsilon  \;.
\end{equation}
Substituting this and the value of $Q$ from \eref{EquivLQG} into \eref{FeedbackLQG} we arrive at 
\begin{equation}
\label{LQGOptCtrl}
	K = \frac{k\epsilon}{\tmix^\star} \; {\rm I}_{2n}  \;,
\end{equation}
which is exactly \eref{FeedbackLG}.

We can gain a better understanding of this equivalence between the control of LG and LQG systems by noting that within the formalism of LQG control one has what is known as the cheap-control limit~\cite{WM10,Fra79}. This is when the cost of control is so cheap that we can spend as much resource or energy on the control input as we please. This notion is thus effected by taking the limit $Q \to 0$. The Ricatti equation \eref{RicattiEqnForY} can then be approximated by
\begin{equation}
\label{CheapControlP}
	P = Y Q^{-1} Y  \;.
\end{equation}
This is exactly \eref{ApproxEqnForY} for the $P$ and $Q$ given in \eref{EquivLQG}, so we see that performing feedback which is linear in $\hxc$ for LG systems is equivalent to doing optimal LQG control in the cheap-control limit.

\section*{References}

\bibliographystyle{unsrt}
\bibliography{mybib}

\end{document}